\documentclass{emulateapj_rj}

\shorttitle{Wavelength Dependence of Galaxy Structure}

\begin{document}

\title{Dependence of Galaxy Structure on Rest-frame Wavelength and
Galaxy Type\footnote{
Based on observations made with the NASA/ESA Hubble Space Telescope,
obtained at the Space Telescope Science Institute,
which is operated by the Association of Universities for Research in
Astronomy, Inc., under NASA contract NAS 5-26555. These observations
are associated with HST programs \#8645, and 9124.}}

\author{Violet A. Taylor-Mager\altaffilmark{1,2} \email{vmager@ociw.edu}}
\author{Christopher J. Conselice\altaffilmark{4}}
\author{Rogier A. Windhorst\altaffilmark{3}}
\author{Rolf A. Jansen\altaffilmark{3}}

\altaffiltext{1}{Department of Physics and Astronomy, Arizona State 
University, Tempe, AZ 85287-1504}
\altaffiltext{2}{Carnegie Observatories, 813 Santa Barbara St., Pasadena,
CA 91101}
\altaffiltext{3}{School of Earth and Space Exploration, Arizona State
University, Tempe, AZ 85287-1404}
\altaffiltext{4}{School of Physics and Astronomy, University of Nottingham,
University Park, England NG7 2RD, UK}

\begin{abstract}

We present a quantitative analysis of the morphologies for 199 nearby 
galaxies as parameterized with measurements of the concentration, asymmetry, 
and clumpiness (CAS) parameters at wavelengths from 0.15-0.85$\mu$m.
We find that these CAS parameters depend on both galaxy type and the 
wavelength of observation. As such, we use them to obtain
a quantitative measure of the "morphological k-correction", i.e., the 
change in appearance of a galaxy with rest-frame wavelength. Whereas 
early-type galaxies (E--S0) appear about the same at all wavelengths 
longward of the Balmer break, there is a mild but significantly-determined 
wavelength-dependence of the CAS parameters for galaxies types later than 
S0, which generally become less concentrated, and more asymmetric and 
clumpy toward shorter wavelengths. Also, as a merger progresses from 
pre-merger via major-merger to merger-remnant stages, it evolves through 
the CAS parameter space, becoming first less concentrated and more 
asymmetric and clumpy, and then returning towards the "locus" of normal 
galaxies. The final merger products are, on average, much more 
concentrated than normal spiral galaxies. 

\end{abstract}


\section{Introduction}

Deep, high-resolution imaging with telescopes such as the Hubble Space 
Telescope (\emph{HST}) 
has revealed that the majority of high redshift galaxies are 
morphologically different from the majority of galaxies nearby.
They appear to have morphologies similar to the 
relatively rare, low redshift irregular and peculiar galaxies 
(e.g., Driver et al. 1995, 1998; Glazebrook et al. 1995; 
Abraham et al. 1996; Odewahn et al. 1996; Ellis 1997; Im et al. 1999; 
van Dokkum et al. 1999; van den Bergh et al. 2000; Conselice
et al. 2005; Papovich et al. 2005). The greater percentage 
of merging galaxies at high redshift suggests that
there was an evolution in the number of merging and interacting galaxies 
throughout the history of the Universe, which supports models of  
hierarchical galaxy formation (e.g., White 1979; White \& Frenk 1991; 
Cole et al. 1994; Kauffmann et al. 1997; Roukema et al. 1997; 
Baugh et al. 1998; Nagashima et al. 2001, 2002). To reveal how galaxies 
assemble and evolve over time,
it is therefore essential to conduct detailed comparisons of galaxies 
as a function of redshift. To carry out these comparisons it is
critical to have a representative baseline at $z \sim 0$.  We present
in this paper a sample of 199 nearby galaxies, observed from the
ultraviolet to the optical red, which can be used as an effective comparison
to higher redshift sources.

Morphology has also arisen in recent years as a major method for 
findings major mergers between galaxies and understanding their
role in galaxy formation (Conselice 2006).  Mergers are predicted
to be the dominate factor in galaxy formation within the Cold
Dark Matter model (e.g., Cole et al. 2000), and testing this concept
in detail is of major importance.  
Extensive work has gone into improving on the traditional Hubble sequence,
which does not distinguish between the large variety of galaxies 
that fall into the catch-all irregular class (e.g. irregular, peculiar 
and merging galaxies). It is of particular importance,
especially at high redshift, to find a method that quantifies the morphologies
of both normal galaxies and the dominant population of irregular/peculiar 
galaxies. Hubble typing is also a
somewhat subjective method of classification that requires visual 
inspection of each individual galaxy. With the advent of large deep sky 
surveys that produce images of hundreds to thousands of galaxies at a time, 
it is essential to develop an automated method of objectively 
classifying galaxies, using a system that describes something about the
physical properties that shape their light distributions, independent
of rest-frame wavelength.

One of the parameters developed for use in automated galaxy classification 
is the concentration index (C) (Abraham et al. 1994, 1996). This parameter
measures how centrally concentrated the light distribution is within
a galaxy, which is a tracer of the disk-to-bulge ratio. Abraham et al.
(1994, 1996) show that classifications through this parameter are less
dependent on high spatial resolution than on Hubble type, and thus 
are more robust at high redshift and in poor ground-based seeing conditions. 
Schade et al. (1995), Abraham et al. (1996) and Conselice (1997) introduced
very basic methods for measuring the rotational asymmetry index (A) to
describe the degree of peculiarity and asymmetry in the light
distribution of nearby galaxies and galaxies imaged with HST at 
redshifts of z $\sim 0.5-1.2$.  The asymmetry index was later
vastly improved through the use of standardized radii and centering
techniques (Conselice et al. 2000).
The asymmetry parameter is particularly useful for identifying merging and 
strongly interacting galaxies, which tend to have very high asymmetries. 
Another parameter, introduced by Isserstedt \& Schindler
(1986), quantifies the ratio of smoothly distributed light to the
clumped light distribution. This clumpiness index (S) compares the
amount of light in star-forming clusters and associations to the light
in a more diffuse older disk population, which correlates with hydrogen
recombination lines (H$\alpha$) (Takamiya 1999), and gives an 
indication of the recent star-formation activity within the galaxy
(Conselice 2003).

Other authors proposed improved methods of measuring these concentration,
asymmetry, and clumpiness (CAS) parameters, making them less sensitive 
to the choice of galaxy center, surface-brightness cut-offs, signal-to-noise 
(S/N) ratio, 
and resolution effects (e.g. Bershady et al. 2000; Conselice et al. 
2000; Conselice 2003; Lotz et al. 2004). The CAS parameters were found to 
correlate with each other,
as well as other fundamental parameters such as galaxy color, luminosity, 
size, surface-brightness, and star formation rate 
(Isserstedt \& Schindler 1986; Abraham et al. 1996; Takamiya 1999; 
Bershady et al. 2000; Corbin et al. 2001). Galaxy types 
are separated-out within parameter spaces involving color, 
surface-brightness, and various combinations of the CAS parameters, 
providing a method of galaxy classification that is relatively robust over a 
large range of redshifts (Bershady et al. 2000; Conselice et al. 2000, 
Conselice et al. 2003;
Conselice 2003). 

Galaxies, however, can look substantially different at shorter
wavelengths than at longer ones (e.g., Bohlin et al. 1991; 
Hill et al. 1992; Kuchinski et al. 2000, 2001; Marcum et al. 2001; 
Windhorst et al. 2002). Hence, if the morphological
$k$-correction is significant, CAS measurements for low redshift
galaxies in rest-frame optical light and high
redshift galaxies in rest-frame ultraviolet light may not be directly 
comparable. For 
example, Schade et al. (1995) found that their galaxies (z $\sim 0.5-1.2$) 
looked more irregular in their $B$ images (rest-frame UV) than at longer 
wavelengths, which will all have an effect on quantitative galaxy parameters. 
Jansen (2000) noted a similar systematic shift from $U$ to $R$ for
galaxies in the Nearby Field Galaxy Survey. We also
observe this effect when comparing HST/WFPC2 UV and near-infrared (IR) 
images available
for a subset of the galaxies presented in Windhorst et al. (2002).
In the present paper, we quantitatively analyze how the CAS 
parameters vary as a function of rest-frame wavelength and morphological type
for a sample of 199 nearby galaxies observed in a combination 
of nine pass-bands ranging from the far-UV to the near-IR. We will
apply these results to the high redshift Universe in a future paper 
(Conselice, et al. 2007, in preparation).

\section{Observations}

\begin{figure*}
\plotone{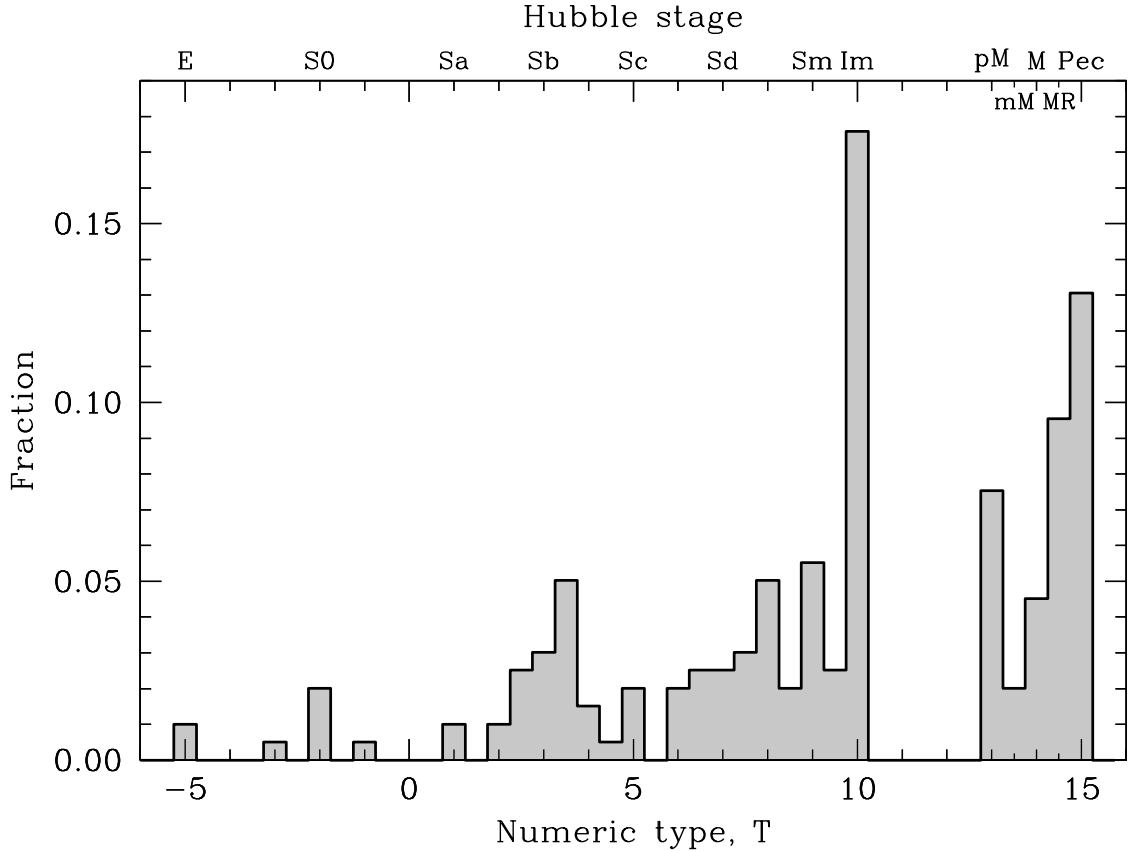}
\figcaption[f1.eps]{{Distribution of types within the present
sample of nearby galaxies. Since the vast majority of galaxies observed
at high redshift display morphologies resembling those of nearby
late-type spiral, irregular, and peculiar/merging galaxies, our sample
is intentionally heavily weighted toward such galaxy types.
The subdivision of types later than 11 is discussed in the text, and
in Hibbard et al. (2001).}}
\end{figure*}

As described in detail in Taylor et al. (2005), we have selected
for study a sample of 199 nearby galaxies that consists mostly of late-type
spiral, irregular, and peculiar/merging galaxies. These late-types 
are among the galaxies at z $\sim 0$ that most resemble
the majority of high redshift galaxies, and are less
studied and less well-understood than nearby earlier-type galaxies. Also,
Jansen (2000) found that most of the wavelength dependence of the asymmetry
is due to galaxies of type Sbc and later (de Vaucouleurs type, T $\gtrsim 4$). 
Thus, most of the difference in asymmetries between the $U$ and $R$-band 
occurs for later morphological Hubble types. As such, late-type galaxies are 
more suited for this analysis than early-type galaxies. The selected 
morphological distribution for our sample is shown in Figure 1. For a 
discussion of the effects of this deliberate type pre-selection, 
see Windhorst et al. (2002).

Of the 199 galaxies in our sample, 143 were observed 
with the Vatican Advanced Technology Telescope (VATT) in $UBVR$ 
($\lambda_c$ = 3597, 4359, 5395, and 6338$\AA$, respectively). A total of 
90 galaxies were observed with HST WFPC2 in the UV 
(F300W, $\lambda_c=2930$\AA), 34 of which also have VATT data. Of these 
90 galaxies observed with HST, 37 were observed in our
Cycle~9 GO program \#8645 (Windhorst et al.\ 2002; Taylor et al.\ 2005), 
and 53 were observed
in our Cycle~10 SNAP-shot program \#9124 (P.I.: R. Windhorst; Jansen et al.
2007, in preparation). During these programs, a subset of 11 galaxies were
also imaged with a shorter wavelength UV filter (F255W, $\lambda_c=2550$\AA),
and 60 were observed in the red (F814W, $\lambda_c=8230$\AA). 
Taylor et al. (2004, 2005) give a more detailed description of these
observations. At the time the present analysis was completed, 13 of the 
galaxies in our sample had deep (t$_{exp} \gtrsim 670$ s) GALEX mid-UV
(NUV, $\lambda_c=2275$\AA) and far-UV (FUV, $\lambda_c=1550$\AA) imaging 
available. A 14th galaxy, NGC0569, was observed only in the GALEX NUV 
filter. Of these, 4 were observed with our GALEX Cycle~1 SNAP program
\#036, and 10 were obtained from the archived Nearby Galaxies
Survey (NGS), Medium Imaging Survey (MIS), and Deep Imaging Survey (DIS)
(Martin et al. 2005; Heckman et al. 2005).
Seven of the 14 galaxies observed with GALEX also have VATT data, and 10 
also have HST data. Three of these galaxies have data from all of these 
telescopes (VATT, HST, and GALEX).

In Table 1 we identify the galaxies in our sample, and list their relevant
global properties. The first 25 galaxies are listed here, with the
full table available in the electronic edition. Column (1) lists the galaxy 
identification numbers
as used throughout this paper, while column (2) gives the common
catalog names, and columns (3) and (4) contain the equatorial
coordinates (J2000.0). Column (6) and (7) lists the total ($B-V$) color
and their errors (respectively), as
measured from the galaxies for which we have $B$ and $V$ images, or from
the RC3 (de~Vaucouleurs et al. 1991) otherwise. All recessional
velocities with respect to the Galactic Standard of Rest
(V$_{GSR}$; column 8) are from the RC3. In column (9), we
identified the edge-on spiral galaxies in our sample. The numeric types
adopted throughout the present paper are listed in column (5),
and were assigned as described in section 4.1. Figure 1
shows the distribution of morphological types within our sample. Note
the deliberate emphasis on late-type galaxies for the purposes of
high redshift comparison.

%
\begin{deluxetable*}{llrrrrrrr}
\tabletypesize{\scriptsize}
\tablecolumns{9}
\tablewidth{5.0in}
\tablecaption{Observed Galaxy List.}
\tablehead{
\colhead{ID\#} & \colhead{Galaxy} & \colhead{RA}
& \colhead{DEC} & \colhead{Type} & \colhead{B-V} & \colhead{$\sigma_{B-V}$}
& \colhead{V$_{GSR}$} & \colhead{comment} \\
\colhead{(1)} & \colhead{(2)} & \colhead{(3)} & \colhead{(4)} &
\colhead{(5)} & \colhead{(6)} & \colhead{(7)} & \colhead{(8)} &
\colhead{(9)}
}
\startdata
 001 & UGC00006 &    00:03:09.46 & +21:57:37.6 & 15.0 &  0.370 & 0.040 &  6763
  & \nodata \\
 002 & NGC0014 &     00:08:46.32 & +15:48:56.4 & 10.0 &  0.580 & 0.035 &  1012
  & \nodata \\
 003 & UGC00156 &    00:16:46.30 & +12:21:13.1 &  9.3 &  0.576 & 0.042 &  1267
  & \nodata \\
 004 & NGC0178 &     00:39:08.25 & -14:10:20.7 &  9.5 &  0.470 & 0.040 &  1496
  & \nodata \\
 005 & UGC00404 &    00:39:19.17 & +13:06:40.3 & 13.0 &  0.362 & 0.030 & 10742
  & \nodata \\
 006 & Mrk960 &      00:48:35.44 & -12:42:59.9 &  8.0 & \nodata & 0.024 &  6447
  & \nodata \\
 007 & UGC00512 &    00:50:02.59 & +07:54:55.3 & 15.0 & -0.004 & 0.017 &  5496
  & \nodata \\
 008 & UGC00644 &    01:03:16.65 & +14:02:01.6 & 13.0 &  0.662 & 0.013 &
 \nodata & \nodata \\
 009 & UGC00685 &    01:07:22.29 & +16:41:04.1 & 10.0 & \nodata & 0.010 &   271
  & \nodata \\
 010 & UGC00749 &    01:11:30.28 & +01:19:10.9 & 14.5 &  0.653 & 0.006 &  6882
  & \nodata \\
 011 & NGC0428 &     01:12:55.62 & +00:58:52.2 &  9.0 &  0.440 & 0.014 &  1231
  & \nodata \\
 012 & UGC00849 &    01:19:23.03 & +12:26:57.4 & 14.5 &  0.572 & 0.015 & 14410
  & \nodata \\
 013 & NGC0569 &     01:29:07.16 & +11:07:53.3 & 13.0 & \nodata & 0.013 &  5862
  & \nodata \\
 014 & UGC01104 &    01:32:43.47 & +18:19:01.4 & 10.0 &  0.533 & 0.018 &   775
  & \nodata \\
 015 & UGC01133 &    01:35:00.85 & +04:23:11.8 & 15.0 &  0.077 & 0.014 &  2031
  & \nodata \\
 016 & NGC0625 &     01:35:05.12 & -41:26:08.9 & 10.0 &  0.560 & 0.016 &   312
  & \nodata \\
 017 & UGC01219 &    01:44:20.13 & +17:28:42.9 &  3.0 &  0.709 & 0.015 &  4708
  & \nodata \\
 018 & UGC01240 &    01:46:19.56 & +04:15:52.5 & 15.0 &  0.401 & 0.015 &  1862
  & \nodata \\
 019 & UGC01449 &    01:58:04.15 & +03:05:57.2 & 14.0 &  0.487 & 0.015 &  5590
  & \nodata \\
 020 & UGC01753 &    02:16:34.98 & +28:12:16.1 &  8.7 &  0.559 & 0.038 &  3098
  & \nodata \\
 021 & NGC0959 &     02:32:23.45 & +35:29:20.1 &  7.0 &  0.588 & 0.009 &   715
  & \nodata \\
 022 & NGC1140 &     02:54:33.43 & -10:01:42.4 & 15.0 &  0.350 & 0.008 &  1479
  & \nodata \\
 023 & NGC1156 &     02:59:41.41 & +25:13:37.5 & 10.0 &  0.618 & 0.009 &   452
  & \nodata \\
 024 & NGC1311 &     03:20:06.66 & -52:11:12.5 & 10.0 &  0.460 & 0.021 &   331
  & \nodata \\
 025 & ESO418-G008 &  03:31:30.58 & -30:12:46.6 &  8.0 &  0.410 & 0.012 &  1146
  & \nodata \\
\enddata
\tablecomments{{\bf Columns:} (1) ID number assigned to this galaxy, (2)
galaxy name, (3) Right Ascension (J2000), (4) Declination (J2000),
(5) classification (de Vaucouleurs numerical types were used for normal
galaxies, with the following types assigned to peculiar/merging galaxies
(see text for details):
13.0=pre-merger, 13.5=minor merger,14.0=major merger,14.5=merger remnant,
15.0=peculiar), (6) total ($B-V$) color (mag), (7) uncertainty on ($B-V$),
(8) Galactic standard of
rest velocity (km s$^{-1}$), and (9) special comment for this galaxy.
"Edge-on" galaxies appear to be edge-on spiral galaxies.
The full table is available only in the electronic edition.
}
\end{deluxetable*}

\section{Data Reduction}

The $UBVR$ CCD images obtained at the VATT were reduced and calibrated 
as described in Taylor et al. (2004, 2005). For the data analysis in 
these previous papers, all of the 
non-target objects in the images were replaced with a constant local 
sky-level. This was suitable for computing radial profiles of
surface-brightness and color, which are not sensitive to structure on
small scales within a galaxy. However, the deviations from the true
underlying galaxy light distribution at the locations 
of non-target objects, as replaced with the local sky-level,
are larger than is acceptable for CAS measurements, which are particularly 
sensitive to high spatial frequency structure within the galaxy.
Therefore, in these cases we applied a more
sophisticated means of removing the non-target objects, interpolating 
from adjacent pixels with valid data over these masked pixels along both
the vertical and horizontal directions (implemented by R. Jansen as task 
\texttt{IMCLEAN}\footnote{http://www.public.asu.edu/$\sim$rjansen/} within 
IRAF\footnote{IRAF is distributed by the National 
Optical Astronomy Observatories, which are operated by the Association 
of Universities for Research in Astronomy, Inc., under cooperative 
agreement with the National Science Foundation.}). This routine
preserves the surrounding sky properties as much as possible inside
the interpolated region. 

We obtained stacked images for most of our HST Cycle~9 galaxies as type 
B associations from the Space Telescope Science Institute (STSCI) 
data archive.\footnote{http://archive.stsci.edu/hst/wfpc2/index.html}  
Seven of the Cycle~9 images do not have associations available for
at least one of the filters. These individual images
were thus combined and cosmic-ray rejected using an IDL routine
developed by Cohen et al. (2003). The HST Cycle~10 images were combined 
using an IRAF task developed by R. Jansen, that rejects bad pixels based 
on the noise characteristics of the images (Jansen et al. 2007, 
in preparation). For each galaxy, the stacked images for the individual
WFPC2 CCD's were then mosaiced using the \texttt{WMOSAIC} task 
within the \texttt{STSDAS} package in IRAF. Non-target
objects were interpolated-over with \texttt{IMCLEAN}.

We retrieved pipeline processed images of each of the sample galaxies
observed by GALEX through the STSCI data archive.
Since GALEX has a field-of-view of $1.2\degr$, and most of our galaxies 
are only around $1\arcmin$ in size, we cut $512\times512$ pixel GALEX 
stamps centered
on our target galaxy for use in the data analysis. This corresponds to
an image size of about $12\arcmin$, which is sufficient for viewing the
entire galaxy and measuring the sky from surrounding regions that are 
far enough 
away from the galaxy that they are uncontaminated by its light. Non-target
objects were interpolated-over with \texttt{IMCLEAN}, as before.

\section{Data Analysis}

\subsection{Visual Classifications}

We assigned visual types to these galaxies using
the average of three experienced classifiers to classify the galaxies
observed with the VATT (Taylor et al. 2005), and the average of
two experienced classifiers (V. Taylor, and C. Conselice) for the galaxies
observed with HST. For each of the normal galaxies in our sample, we 
assigned morphological types on the numeric 16-step de Vaucouleurs system,
which ranges from T$=-5$ (Elliptical) through T$=10$ (irregular).
The median difference between classifications by
individual classifiers was zero, with 75\% of the classifications 
agreeing to within $\pm 1$ T-type-bin, and all agreeing to within $\pm 4$
bins. For merging and peculiar galaxies, we used the following typing scheme,
which was partially adapted from Hibbard et al. (2001). Pre-mergers
(pM, T$=13.0$) are galaxies that are tidally interacting with another
nearby galaxy, but they are at a large enough distance from each other
that the individual galaxies can be easily distinguished and treated
separately. Minor mergers (mM, T$=13.5$) are two or more galaxies
showing signs of merging, in which one of the galaxies is much
larger or brighter than the others (estimated to be at least about 
4 times larger or 
brighter). Major mergers (M, T$=14.0$) are galaxies of apparently similar
mass or luminosity that are interacting or merging. Major merger remnants 
(MR, T$=14.5$) are objects in the later stages of merging, 
such that it is difficult 
to say exactly what has merged, when, or how. Peculiar galaxies
(P, T$=15.0$) are abnormal or unclassifiable galaxies which do not
fit on the normal Hubble sequence, but were not obviously involved in
a merger. A subset of these may, however, be late-stage merger remnants.
Figure 2 shows an example of each of the four major types of 
merging galaxies.

\begin{figure*}
\plotone{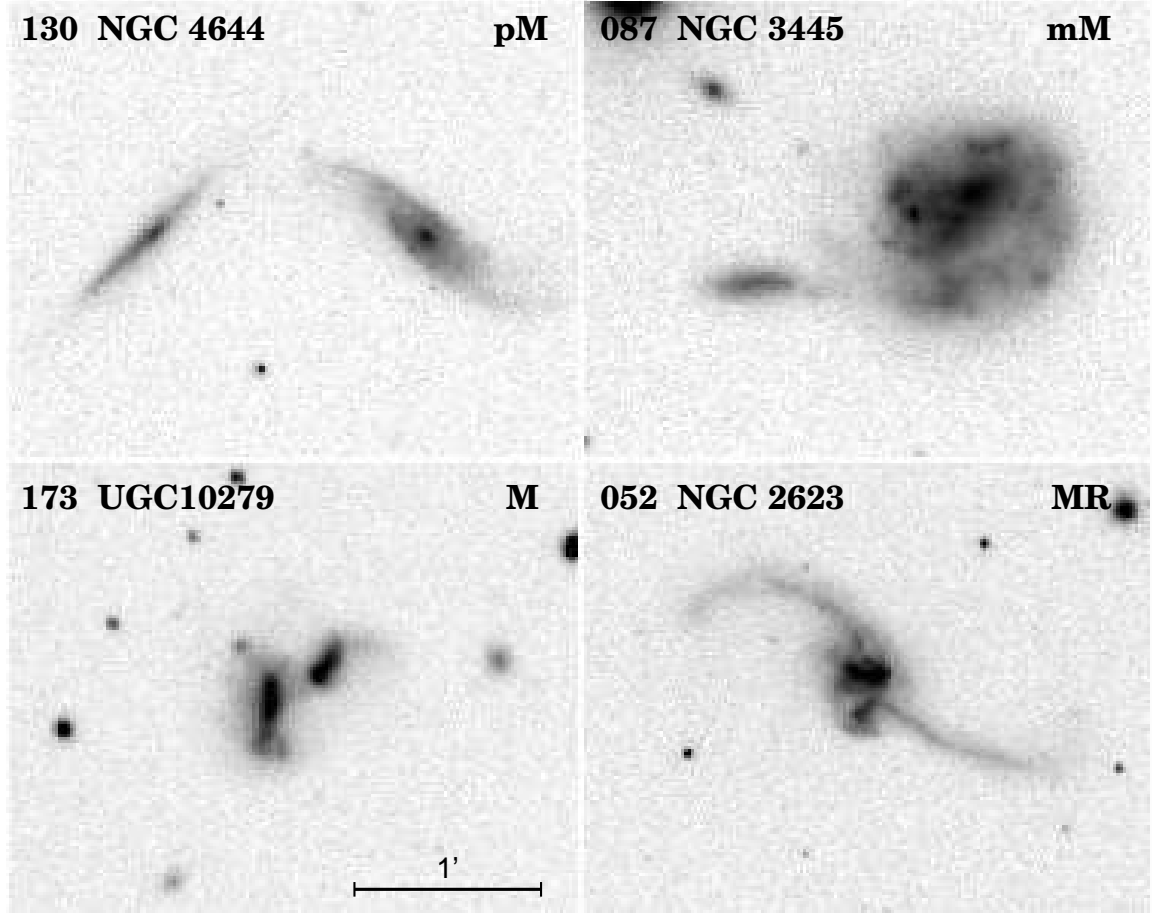}
\figcaption[f2.eps]{{Examples of the sub-classifications assigned
to the merging galaxies in our sample. Each of the the B-filter VATT images
measures $3.0\arcmin \times 2.4\arcmin$. {\bf Top left:} NGC4644 (the galaxy
on the right) is in strong interaction with its neighbor NGC4644B (PGC42725),
and is classified as a pre-merger (pM). {\bf Top right:} NGC3445 appears to be
accreting a smaller companion, PGC32784, and is classified as a minor
merger (mM). {\bf Bottom left:} UGC10279 represents a major merger event (M)
between two galaxies of similar mass, Holmberg 734A and B. {\bf Bottom right:}
in NGC2623 the properties of the galaxies that merged are no longer exactly
discernible, so it is classified as a merger remnant (MR, early stage).
}}
\end{figure*}

\subsection{CAS Parameter Measurements}

We adopt the definitions of the CAS parameters of Conselice et al. 
(2000) and Conselice (2003), as described below, and measured CAS
parameters for our galaxy sample using the IRAF task \texttt{CAS}, 
developed by C. Conselice (Conselice et al. 2000; Conselice 2003).

The concentration index, C, is computed by determining the total
sky-subtracted light contained within $1.5 \times$ the Petrosian radius 
(r$_{Pet}$) (Petrosian 1976), 
and finding the logarithmic ratio of
the radius within which 80\% of this light is contained (r$_{80}$) to the 
radius within which 20\% of this light is contained (r$_{20}$)
(Conselice et al. 2000).
The concentration index is thus given by the formula:
\begin{equation}
C = 5 \times log(r_{80}/r_{20})
\end{equation}
Therefore, galaxies that are highly concentrated in their centers
(e.g., ellipticals) will have high values of C, and galaxies that
have less-centrally concentrated light distributions (e.g., disk
galaxies or low surface-brightness galaxies) will have lower values of C.

The asymmetry index, A, is computed by rotating the galaxy by 
180\degr~from its center, and subtracting the
light within $1.5 \times$ r$_{Pet}$ in the rotated image (I$_{180}$) 
from that in the original image (I$_o$). The ratio of the residual 
subtracted flux to the original galaxy flux yields A:
\begin{equation}
A = min ({\Sigma~| I_o - I_{180} | \over {\Sigma~| I_o |}}) -
min ({\Sigma~| B_o - B_{180} | \over {\Sigma~| I_o |}})
\end{equation}
Noise corrections were applied by subtracting the asymmetry of an empty
background region (B), and iterative centering corrections were applied 
to minimize A, since gross centering errors would artificially increase
A. Asymmetries range from A = 0 to 2, with A = 0 corresponding to a truly 
symmetric galaxy (e.g., some ellipticals; see Jansen(2000) for an example
of a highly symmetric spiral galaxy), and A = 2 corresponding to
a completely asymmetric galaxy.

The clumpiness index, S, is defined as the ratio of the amount of light
in high spatial frequency structures within $1.5 \times$ r$_{Pet}$ to the 
total amount of light in the galaxy within that radius (Conselice 2003).
This was done by subtracting a boxcar-smoothed image from the original image
to produce an image that contains only the high-frequency structure.
The central pixels within 1/20th of the total radius were set to zero
to exclude them from the parameter measurements.
S is given by the following equation, where I$_{x,y}$ is the intensity 
of light in a given pixel, I$^s_{x,y}$ is the intensity of 
that pixel in the image smoothed by $0.3 \times $r$_{Pet}$, 
and B$^s_{x,y}$ is an intensity 
value of a pixel from a smoothed background region: 
\begin{equation}
S = 10 \times ({\Sigma^{N,N}_{x,y=1,1}~(I_{x,y} - I^s_{x,y}) \over 
\Sigma^{N,N}_{x,y=1,1}~I_{x,y}} - 
{\Sigma^{N,N}_{x,y=1,1}~B^s_{x,y} \over \Sigma^{N,N}_{x,y=1,1}~I_{x,y}})
\end{equation}
A clumpiness of S = 0 corresponds to a galaxy that has no high frequency
structure, and is therefore completely smooth (e.g., some ellipticals that
are far enough away that they are not resolved into their brightest stars).
Galaxies with more high-frequency structure are more patchy in appearance,
and will have a higher value of S. 

\section{Resolution and Signal-to-Noise Effects}

The effects of spatial resolution on the CAS parameters are discussed
in detail in Conselice (2003) and Conselice et al. (2000).
Previous studies have shown that a spatial resolution (R) of $\sim 1$~kpc 
is at the limit beyond which poorer resolution will affect the measurements, 
with asymmetry parameters determined with HST/WFPC2 becoming unreliable 
for the more distant galaxies in the Hubble Deep Field 
(Conselice et al. 2000; Lotz et al. 2004). At resolutions better than this 
limit, Conselice (2003) found that the asymmetry index does not depend 
strongly on the resolution or seeing of the image, with 70\% of the value of
A, as measured in the optical, attributable to overall global distortions 
in the galaxy, and 30\% due to localized high-frequency structures, such as 
star-forming regions. Low S/N ratios
within the galaxy can also decrease the reliability of the
CAS measurements (Conselice et al. 2000; Lotz et al. 2004). Conselice et al.
(2000) found that asymmetries can be computed most reliably when the 
integrated S/N ratio $\gtrsim 100$ per object.
These S/N and resolution reliability limits, however, are rather 
conservative, and would result in poor (or no) number statistics in
some of our galaxy type-bins. Therefore, in this section we will 
investigate the effects of S/N and spatial resolution in our data, 
and determine the limits of each that are appropriate for our analysis. 
We determined the 
spatial resolution in kpc from the recessional velocities listed in the
RC3 (using V$_{GSR}$ when available, or V$_{hel}$ otherwise).
Total S/N ratios were measured within 1.5 times the Petrosian
radius. 

\subsection{The Effects of Smoothing the HST data}

For the $UBVR$ images observed at the VATT, the images of each galaxy were 
convolved to the worst seeing for that galaxy (typically in the $U$-band), 
using a Gaussian of the appropriate width within the task \texttt{GAUSS} in 
IRAF. The seeing of each image was determined
by finding the average FWHM (full width at half max) of all stars detected 
with SExtractor (Bertin \& Arnouts 1996), as described in detail in 
Taylor et al. (2004). 

In order to consistently compare images of a galaxy with ground-based seeing 
to those from HST/WFPC2 (FWHM$\sim$0\farcs14), it may be necessary
to convolve the HST images to the same resolution as the ground-based
data. We test this by examining the effect on the CAS parameter measurements
after convolving the HST images (as for the VATT) to the $1\farcs75$ average 
U-band seeing at the VATT. 
The left three panels in Figure 3 show the difference between the CAS 
measurements from the convolved and unconvolved HST images in each filter 
vs. the measurements in the unconvolved images. The right-most panels show 
the median of these differences for each filter. Horizontal dotted lines 
indicate when convolution does not affect the CAS parameters. 
Different symbols in Figure 3 are used for measurements from
images where the galaxy had low S/N (large open squares: S/N $< 75$,
small open squares: $75 <$ S/N $< 100$), and where the galaxy had low physical
resolution (large open triangles: R $> 1.25$ kpc, small open 
triangles: 1.00 $<$ R $< 1.25$ kpc). There is no real trend with
low S/N or spatial resolution, although most of the extreme outliers
have low S/N or, in exceptional cases, low resolution. 

\begin{figure*}
\plotone{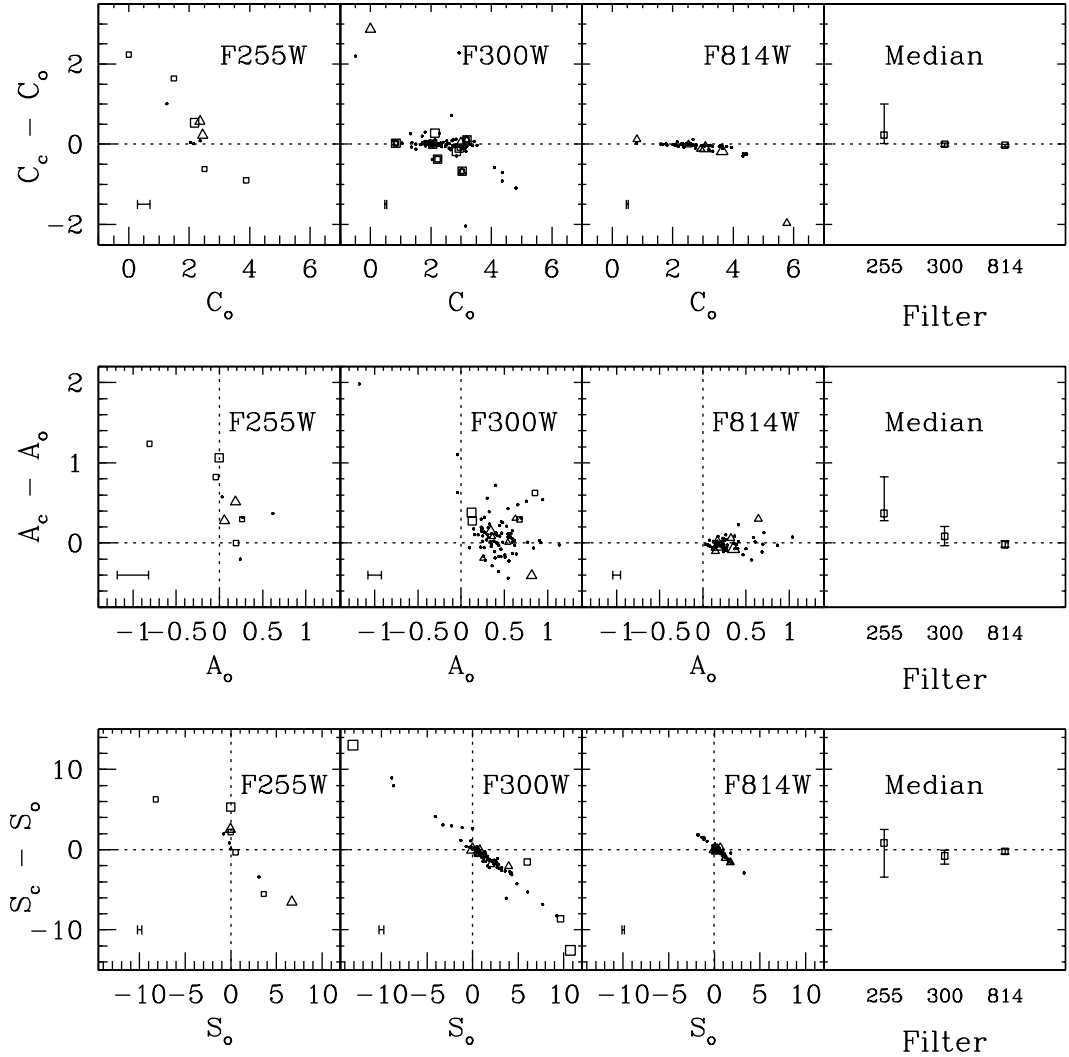}
\figcaption[f3.eps]{{The difference between CAS parameters measured
from HST images convolved to the average VATT ground-based seeing of
$1.75\arcsec$ (X$_c$), and those measured from the original high resolution
($0.14\arcsec$ FWHM) HST images (X$_o$), plotted versus
the measurements in the original images. {\it Open square symbols} represent
galaxies with low S/N ({\it large symbols}: S/N $< 75$,
{\it small}: $75 < $ S/N $< 100$). {\it Open triangle symbols} represent
galaxies with low spatial resolution (R) in
the convolved images ({\it large}: R $> 1.25$ kpc, {\it small}:
1.00 kpc $< R <$ 1.25 kpc).
{\bf Left 3 panels:} {\it TOP:} Change in concentration index,
C$_c -$ C$_o$, versus C$_o$. {\it MIDDLE:} Change in asymmetry index,
A$_c -$ A$_o$, versus A$_o$. {\it BOTTOM:} Change in clumpiness index,
S$_c -$ S$_o$, versus S$_o$. Each panel shows data from a different
pass-band, as labeled in the panels. The horizontal error bar in the lower
left corner of each panel represents the median uncertainties for
each parameter in the original images.
{\bf Right-most panels:} Median difference in each filter between
C$_c$ and C$_o$ (top), A$_c$ and A$_o$ (middle), and
S$_c$ and S$_o$ (bottom). Error bars represent the 25\% -- 75\%
quartile range. Horizontal dotted lines indicate where the value of the
CAS parameter did not change between the convolved and unconvolved images.
Vertical dotted lines in the A and S panels represent the cut-off below
which the A and S measurements of the sky were larger than those of
the galaxy. Most of these outlying values were increased to positive
values after convolving the images.
}}
\end{figure*}

The differences between CAS parameters at the two different resolutions
are most pronounced in the F255W images, which had the lowest 
S/N ratio, and thus the highest uncertainties in these
parameters. Convolving these images resulted in boosting the signal in
each resolution element, as well as smoothing over 
high-frequency structure in the noisy background, which is strongly 
affected by charge transfer effects (CTE). This
results in more reliable CAS values that are not as severely affected 
by spikes in the sky background, as can be seen by considering the 
A and S panels in Figure 3.
Vertical dotted lines in these panels represent the limit below which
the A and S values measured in the sky were higher than those measured
in the galaxy (A $< 0$, S $< 0$). For these extreme cases, convolving 
the images increased the signal and decreased the noise, which resulted in 
increasing most of the negative values of A and S into the positive 
regime. CTE-noise effects are also present in the data from the other 
WFPC2 filters, although it is not as strong of a factor in
the longer wavelength pass-bands due to the higher S/N 
ratio of the galaxies and the more significant zodiacal sky-brightness 
filling in the low-level CTE-traps. A combination of smoothing over the 
noise and the 
high-frequency structure of the galaxy itself has an effect on the 
CAS parameters in all filters, to varying extents. 

Overall, the concentration index does not change significantly with 
spatial resolution, especially at longer wavelengths. The
median values of the change in C after convolving the HST images 
to the ground-based seeing are 
0.228, --0.002, and --0.028 in F255W, F300W, and F814W, respectively (these
values will be given in this filter order for the rest of this paragraph). 
This is marginally greater than the median uncertainty on the
individual unconvolved values of C in F255W (0.209), and 
well within the median uncertainties in F300W and F814W (0.030 for both).
The asymmetry index is more affected by convolution,
although there is no trend in this effect with increasing A. This is
likely due to a boost in S/N per resolution element, which improves
the reliability of the measurement (Lotz et al. 2004). Convolving
the images to the ground-based seeing changed A by a median value of 
0.369, 0.083, and --0.021, respectively. 
Except for the noisy F255W images, this is roughly within the median 
uncertainties of 0.181, 0.078, and 0.044, respectively. The clumpiness 
index, however,
was more strongly affected by the resolution change, with intrinsically
clumpy galaxies showing a strong trend in all filters toward decreasing 
clumpiness with decreasing resolution. This is to be expected, as
S is designed to measure high-frequency structure, and star clusters
and associations may be resolved in the unconvolved HST images. 
Convolving the images changed the median S by 0.840, --0.780, and --0.215,
respectively. 
These are relatively significant differences, and much larger than the
median uncertainties on S of 0.249, 0.256, and 0.135, illustrating that
real structure being smoothed over lowers the measured S values.

With the above caveats, throughout the remainder of this paper we 
therefore use the CAS values measured
in HST images convolved to match the resolution of our ground-based
VATT images, to reduce the effects of background noise and to offer a more
systematic comparison. We note that with the exception of F814W, this
does not do gross injustice to the high resolution HST images, since
the F255W and F300W filters are below the atmospheric cut-off, and therefore
are unique data even at lower resolution.

\subsection{Determining a S/N and Resolution Reliability limit}

Of the 143 galaxies observed with the VATT, 16 have spatial 
resolutions between 1.00 and 2.20 kpc in the U-band image, which is above 
the reliability limit of 1 kpc quoted in Conselice et al. (2000).
Only one galaxy (UGC01133 in the V-band) was found to have a S/N ratio
below the Conselice et al. (2000) reliability limit of 100. These limits
(hereafter referred to as limit$_1$) are
conservative, however, so it may be acceptable to use more relaxed
limits in the interest of conserving higher number statistics. If we relax
the limits by 25\% (defined as limit$_2$), we find 9 galaxies with 
R $> 1.25$ kpc, and no galaxies with S/N $< 75$.  

Of the 90 galaxies observed with the HST in F300W (11 of which were also 
observed in F255W, and 60 in F814W), 6 have $1.00 < $ R $< 1.44$ kpc,
and 2 have R $> 1.25$ kpc. In F255W, 7 galaxies have S/N $< 100$, and 3
have S/N $< 75$. In F300W, 5 galaxies have S/N $< 100$, and 2 have S/N $< 75$.
All galaxies observed in F814W have S/N $> 100$. 

The GALEX images have a stellar FWHM of $\sim$5\arcsec, which is
$\sim$3$\times$ larger than that of the VATT and the convolved HST data. 
The CAS values, however, should still be reliably determined, as long as 
the spatial resolutions of the galaxies are within the above reliability 
limit. Of the 14 galaxies
observed with GALEX, 7 have R $> 1.00$ kpc, with resolutions 
between 1.41 and 2.37 kpc. One additional galaxy (NGC1396) was barely
detected in the FUV. 

The main goal of this work is to determine the general trends of the
CAS parameters, mainly through the use of the median values for each
type-bin and filter. For well-populated bins of type and filter, 
imposing conservative reliability limits improves the significance of 
the median values. In several of our sparsely populated bins, however, 
measurement uncertainties and peculiarities of individual galaxies 
would dominate. If we can
relax the Conselice et al. (2000) limits of S/N $> 100$ and R $< 1.00$ kpc
(limit$_1$) without having a significant effect on the median CAS values,
we can then improve the number statistics in our type-bins while still
obtaining reasonably high reliability. As such, we have tested the 
effects of applying limit$_1$ to our relatively well-sampled VATT and 
HST F300W and F814W data set against applying a more liberal limit of 
S/N $> 75$ and R $< 1.25$ kpc (limit$_2$). Table 2 summarizes the results. 
The median CAS parameters in each type-bin were calculated for
the entire data set ("all"), and separately for those data meeting each of the
reliability requirements we are considering (limit$_1$ and limit$_2$). Since 
we find little difference in the results between filters for each telescope, 
we average the medians over $U$, $B$, $V$, and $R$ for the VATT, and over 
F300W and F814W for HST. Table 2
lists the difference between these medians calculated with limit$_1$ and
all of the data, and also the difference between the medians calculated
with limit$_1$ and limit$_2$. The final column in Table 2 gives the
average uncertainty on the difference between the medians for a given
parameter observed with the VATT or HST. These uncertainties were
calculated by taking the quadratic sum of the median range of CAS values,
averaged over the relevant filters, and normalized by the square-root 
of the average number 
of galaxies per type-bin. Although in a few individual sparsely populated 
type-bins the median CAS values are affected at the $\gtrsim$1$\sigma$ level, 
we find that for the vast majority of bins the median values do not change 
by relaxing our limits on minimum S/N and spatial resolution from limit$_1$
to limit$_2$. This is also the case for the GALEX and F255W data.

%
\begin{deluxetable*}{cllrrrrrrrrr}
\tabletypesize{\scriptsize}
\tablecolumns{12}
\tablewidth{5.5in}
\tablecaption{Change in Median CAS Parameters With Reliability Limits}
\tablehead{
\colhead{Parameter} & \colhead{Telescope} & \colhead{Difference}
& \colhead{E--S0} & \colhead{Sa--Sc} & \colhead{Sd--Im}
& \colhead{Pm} & \colhead{mM} & \colhead{M} & \colhead{MR} &
\colhead{P} & \colhead{$\sigma_{med}$} \\
\colhead{(1)} & \colhead{(2)} & \colhead{(3)} & \colhead{(4)} &
\colhead{(5)} & \colhead{(6)} & \colhead{(7)} & \colhead{(8)} &
\colhead{(9)} & \colhead{(10)} & \colhead{(11)} & \colhead{(12)}
}
\startdata
 C & VATT & limit$_1-$all & 0.120 & 0.000 & 0.000 & -0.005 & 0.358 & 0.183 &
 -0.030 & 0.053 & 0.10 \\
 C & VATT & limit$_1-$limit$_2$ & -0.120 & -0.015 & 0.000 & 0.000 & -0.358 &
 0.245 & 0.035 & -0.053 & 0.10 \\
 C & HST &  limit$_1-$all & 0.000 & 0.045 & -0.040 & -0.195 & 0.000 & 0.435 &
 -0.105 & 0.000 & 0.10 \\
 C & HST &  limit$_1-$limit$_2$ & 0.000 & -0.065 & 0.020 & 0.195 & 0.000
& -0.435 & 0.000 & 0.000 & 0.10 \\
 A & VATT & limit$_1-$all & -0.008 & -0.003 & 0.000 & 0.030 & 0.088 & -0.113
  & 0.008 & 0.003 & 0.04 \\
 A & VATT & limit$_1-$limit$_2$ & 0.008 & 0.000 & 0.000 & 0.000 & -0.088 &
0.115 & -0.008 & -0.003 & 0.04 \\
 A & HST &  limit$_1-$all & 0.000 & -0.010 & -0.005 & 0.005 & 0.000 & 0.010
& 0.055 & 0.000 & 0.06 \\
 A & HST &  limit$_1-$limit$_2$ & 0.000 & 0.015 & 0.005 & -0.005 & 0.000 &
-0.010 & 0.000 & 0.000 & 0.06 \\
 S & VATT & limit$_1-$all & 0.003 & 0.018 & 0.000 & 0.055 & 0.098 & -0.273
  & 0.020 & 0.000 & 0.06 \\
 S & VATT & limit$_1-$limit$_2$ & -0.003 & 0.015 & 0.000 & 0.000 & -0.098 &
 0.143 & 0.015 & 0.000 & 0.06 \\
 S & HST &  limit$_1-$all & 0.000 & 0.045 & 0.010 & 0.010 & 0.000 &
-0.160 & 0.140 & 0.000 & 0.25 \\
 S & HST &  limit$_1-$limit$_2$ & 0.000 & -0.015 & -0.005 & -0.010 & 0.000
& 0.160 & 0.000 & 0.000 & 0.25 \\
\enddata
\tablecomments{{\bf Columns:} (1) Measured parameter (C, A, or S). (2)
Mean values were determined within $U$, $B$, $V$, and $R$ (VATT), or F300W and
F814W (HST). (3) Type of difference calculated: "all" is the mean within
the filter set described by column 2 of the median CAS parameter
within each type-bin for all of the data; "limit$_1$" is this median when
galaxies with S/N $< 100$ and resolution $> 1.00$ kpc are rejected;
"limit$_2$" is this median when galaxies with S/N $< 75$ and
resolution $> 1.25$ kpc are rejected. (4--11) Median CAS parameter difference
(as described in column 3) for each galaxy type-bin.
(12) Average error on the median CAS parameter differences.
}
\end{deluxetable*}

Applying limit$_1$ results in an average change with respect to no limit 
in the median C, A, and S 
values, respectively, of $0.051 \pm 0.035$, $0.004 \pm 0.018$, and 
$-0.002 \pm 0.064$. Within the uncertainty on these averages, only the 
concentration index is affected by the application of limit$_1$. 
Applying limit$_2$ results in an average change from the values calculated
with limit$_1$ of $-0.034 \pm 0.035$, $0.002 \pm 0.018$, 
and $0.013 \pm 0.064$, in C, A, and S, respectively. All of these
are within the uncertainties, although only marginally so for C. Therefore,
we find that it is acceptable to apply the somewhat more relaxed reliability
limits of limit$_2$ to our analysis, in order to retain adequate sample sizes. 
In the subsequent plots, data points 
that do not satisfy our S/N and resolution criteria (limit$_2$) will be 
plotted using smaller symbols, but not used to calculate the presented 
trends. Applying this limit results in the rejection of 8 of the 13 
galaxies in FUV, 7/14 in NUV, 3/11 in F255W, 4/90 in F300W, 9/143 in $UBVR$,
and 2/60 in F814W. It is clear that the FUV, NUV, and F255W data are 
barely adequate for the current studies, and only instruments like
WFC3 (to be launched to HST in SM4) can significantly improve on this.

\section{Results}

\subsection{Relating CAS Parameters to Galaxy Color and Type}

Tables 3, 4, and 5 list our C, A, and S measurements and errors, respectively, 
for each galaxy in each pass-band. The first 25 galaxies are listed here,
with the full tables available in the electronic edition. Ellipsis 
indicate that there are no images available for that galaxy in that 
particular pass-band. Footnotes mark measurements for images in which the
galaxy did not meet the S/N or resolution requirements of our adopted 
limits (S/N$< 75$ or R$> 1.25$ kpc).
Figures 4, 5, and 6 show the C, A, and S parameters, respectively,
versus the total ($B-V$) color of each galaxy. Values for each
pass-band are shown in separate panels within each figure, with different 
colored symbols used to designate early-type (E--S0), mid-type (Sa-Sc), 
late-type (Sd--Im), and peculiar/merging galaxies, as indicated by the
legend in Figure 4. Smaller symbols represent galaxies that are below
the reliability limits adopted in Section 5.2 (S/N$< 75$, or R$>1.25$ kpc). 
Representative error bars are shown in 
each panel with the median uncertainty in the CAS parameters of the 
individual data points, and the average ($B-V$) uncertainty of 0.04 mag.
These figures show a trend of galaxies becoming generally 
more concentrated, less asymmetric, and less clumpy with redder 
($B-V$) color. There is a large spread and overlap of these parameters with 
galaxy type, so that galaxy classifications made from these plots can 
only be used as approximate indicators.
Early-type galaxies are most clearly separated from the later galaxy types, 
since they are the reddest, the most concentrated, the least asymmetric, 
and the least clumpy. Later galaxy types are increasingly bluer, 
less concentrated, more
asymmetric, and more clumpy than earlier galaxy types. This result is
in agreement with previous authors (e.g. Takamiya 1999; Bershady et al. 2000; 
Conselice et al. 2000, 2003; Conselice 2003). There is a much higher 
spread in S at shorter wavelengths, with a CAS-$(B-V)$ correlation 
that changes slope with filter, such that blue 
galaxies tend to be more clumpy at shorter wavelengths than they are 
at longer wavelengths, although the statistics in the UV are scarce. 
This logically
results from bluer late-type galaxies containing more recent star
formation, and therefore having more blue knots from recently 
formed star clusters and associations that will increase the 
clumpiness in the bluer filters. The CAS values of peculiar/merging galaxies
vary considerably, but their median values are 
slightly offset from the locus of normal galaxies, with peculiar/merging
galaxies being generally bluer (by 0.03 mag), and slightly 
more concentrated (by $\sim 0.2$) and asymmetric (by $\sim 0.1$). These
type-dependent offsets and their overlap are discussed in more detail
below.

%
\begin{deluxetable*}{lrrrrrrrrrrrrrrrrrr}
\tabletypesize{\footnotesize}
\tablecolumns{19}
\tablewidth{\textwidth}
\tablecaption{Concentration Index (C)}
\tablehead{
\colhead{ID\#} & \colhead{FUV} & \colhead{$\sigma_{FUV}$} & \colhead{NUV} &
\colhead{$\sigma_{NUV}$} & \colhead{F255W} & \colhead{$\sigma_{F255W}$}
& \colhead{F300W} & \colhead{$\sigma_{F300W}$} & \colhead{$U$} &
\colhead{$\sigma_{U}$} & \colhead{$B$} & \colhead{$\sigma_{B}$} &
\colhead{$V$} & \colhead{$\sigma_{V}$} & \colhead{$R$} &
\colhead{$\sigma_{R}$} & \colhead{F814W} & \colhead{$\sigma_{F814W}$}
}
\startdata
 001 &       2.617\tablenotemark{2} &   0.756 &   2.950\tablenotemark{2} &   0.728 & \nodata & \nodata &   3.722
  &   0.102 & \nodata & \nodata & \nodata & \nodata & \nodata & \nodata &
 \nodata & \nodata & \nodata & \nodata \\
 002 &     \nodata & \nodata & \nodata & \nodata & \nodata & \nodata &   2.254
  &   0.023 & \nodata & \nodata & \nodata & \nodata & \nodata & \nodata &
 \nodata & \nodata &   2.346 &   0.022 \\
 003 &     \nodata & \nodata & \nodata & \nodata & \nodata & \nodata & \nodata
  & \nodata &   2.507 &   0.044 &   2.720 &   0.039 &   2.404 &   0.051 &
   2.866 &   0.039 & \nodata & \nodata \\
 004 &     \nodata & \nodata & \nodata & \nodata & \nodata & \nodata &   2.247
  &   0.025 & \nodata & \nodata & \nodata & \nodata & \nodata & \nodata &
 \nodata & \nodata &   2.211 &   0.024 \\
 005 &     \nodata & \nodata & \nodata & \nodata & \nodata & \nodata & \nodata
  & \nodata &   2.488\tablenotemark{2} &   0.126 &   2.669\tablenotemark{2} &   0.115 &   2.581\tablenotemark{2} &   0.122 &
   2.343\tablenotemark{2} &   0.154 & \nodata & \nodata \\
 006 &       2.487\tablenotemark{2} &   0.487 &   2.784\tablenotemark{2} &   0.489 & \nodata & \nodata &   2.704
  &   0.043 & \nodata & \nodata & \nodata & \nodata & \nodata & \nodata &
 \nodata & \nodata &   3.164 &   0.049 \\
 007 &     \nodata & \nodata & \nodata & \nodata & \nodata & \nodata & \nodata
  & \nodata &   2.632 &   0.113 &   2.652 &   0.107 &   2.770 &   0.110 &
   2.767 &   0.125 & \nodata & \nodata \\
 008 &     \nodata & \nodata & \nodata & \nodata & \nodata & \nodata & \nodata
  & \nodata &   2.125\tablenotemark{2} &   0.116 &   2.275\tablenotemark{2} &   0.116 &   2.512\tablenotemark{2} &   0.127 &
   2.851\tablenotemark{2} &   0.122 & \nodata & \nodata \\
 009 &     \nodata & \nodata & \nodata & \nodata & \nodata & \nodata &   1.801
  &   0.018 & \nodata & \nodata & \nodata & \nodata & \nodata & \nodata &
 \nodata & \nodata & \nodata & \nodata \\
 010 &     \nodata & \nodata & \nodata & \nodata & \nodata & \nodata & \nodata
  & \nodata &   4.426 &   0.165 &   4.242 &   0.143 &   4.088 &   0.147 &
   4.195 &   0.140 & \nodata & \nodata \\
 011 &     \nodata & \nodata & \nodata & \nodata & \nodata & \nodata &   1.407
  &   0.015 & \nodata & \nodata & \nodata & \nodata & \nodata & \nodata &
 \nodata & \nodata & \nodata & \nodata \\
 012 &     \nodata & \nodata & \nodata & \nodata & \nodata & \nodata & \nodata
  & \nodata &   2.848\tablenotemark{2} &   0.115 &   2.860\tablenotemark{2} &   0.106 &   2.924\tablenotemark{2} &   0.109 &
   2.937\tablenotemark{2} &   0.113 & \nodata & \nodata \\
 013 &     \nodata & \nodata &   3.150\tablenotemark{2} &   0.378 & \nodata & \nodata &   2.852
  &   0.062 & \nodata & \nodata & \nodata & \nodata & \nodata & \nodata &
 \nodata & \nodata &   3.060 &   0.064 \\
 014 &       2.489 &   0.485 &   2.622 &   0.448 & \nodata & \nodata &   2.510
  &   0.032 &   2.490 &   0.112 &   2.682 &   0.105 &   2.785 &   0.096 &
   3.341 &   0.072 &   2.571 &   0.026 \\
 015 &     \nodata & \nodata & \nodata & \nodata & \nodata & \nodata & \nodata
  & \nodata &   2.373 &   0.062 &   2.442 &   0.064 &   2.537 &   0.080 &
   2.528 &   0.114 & \nodata & \nodata \\
 016 &     \nodata & \nodata & \nodata & \nodata & \nodata & \nodata &   2.958
  &   0.031 & \nodata & \nodata & \nodata & \nodata & \nodata & \nodata &
 \nodata & \nodata & \nodata & \nodata \\
 017 &     \nodata & \nodata & \nodata & \nodata & \nodata & \nodata & \nodata
  & \nodata &   3.022 &   0.096 &   3.399 &   0.113 &   3.816 &   0.135 &
   3.997 &   0.147 & \nodata & \nodata \\
 018 &     \nodata & \nodata & \nodata & \nodata & \nodata & \nodata & \nodata
  & \nodata &   3.101 &   0.127 &   3.041 &   0.127 &   2.941 &   0.133 &
   2.603 &   0.168 & \nodata & \nodata \\
 019 &     \nodata & \nodata & \nodata & \nodata & \nodata & \nodata & \nodata
  & \nodata &   1.602 &   0.058 &   1.665 &   0.057 &   1.629 &   0.057 &
   1.548 &   0.057 & \nodata & \nodata \\
 020 &     \nodata & \nodata & \nodata & \nodata & \nodata & \nodata & \nodata
  & \nodata &   2.526 &   0.106 &   2.458 &   0.106 &   2.595 &   0.100 &
   2.546 &   0.101 & \nodata & \nodata \\
 021 &     \nodata & \nodata & \nodata & \nodata & \nodata & \nodata & \nodata
  & \nodata &   2.133 &   0.044 &   2.250 &   0.042 &   2.363 &   0.042 &
   2.420 &   0.043 & \nodata & \nodata \\
 022 &     \nodata & \nodata & \nodata & \nodata & \nodata & \nodata &   3.654
  &   0.095 & \nodata & \nodata & \nodata & \nodata & \nodata & \nodata &
 \nodata & \nodata &   3.488 &   0.051 \\
 023 &     \nodata & \nodata & \nodata & \nodata & \nodata & \nodata &   1.905
  &   0.018 &   2.760 &   0.034 &   2.745 &   0.032 &   2.718 &   0.031 &
   2.681 &   0.030 &   1.790 &   0.017 \\
 024 &     \nodata & \nodata & \nodata & \nodata & \nodata & \nodata &   1.970
  &   0.021 & \nodata & \nodata & \nodata & \nodata & \nodata & \nodata &
 \nodata & \nodata &   2.046 &   0.020 \\
 025 &     \nodata & \nodata & \nodata & \nodata & \nodata & \nodata &   1.974
  &   0.020 & \nodata & \nodata & \nodata & \nodata & \nodata & \nodata &
 \nodata & \nodata &   2.271 &   0.023 \\
\enddata
\tablecomments{{\bf Columns:} Concentration indices in each filter
and their uncertainties ($\sigma_{filter}$). The ID\# is the
identification number assigned to each galaxy in Table 1.
The full table is available only in the electronic edition.
}
\tablenotetext{1}{Galaxy does not meet the S/N requirements for
accurate CAS parameter measurements in this image (S/N $< 75$).}
\tablenotetext{2}{Galaxy does not meet the resolution requirements for
accurate CAS parameter measurements in this image (R $> 1.25$ kpc).}
\end{deluxetable*}

%
\begin{deluxetable*}{lrrrrrrrrrrrrrrrrrr}
\tabletypesize{\scriptsize}
\tablecolumns{19}
\tablewidth{\textwidth}
\tablecaption{Asymmetry Index (A)}
\tablehead{
\colhead{ID\#} & \colhead{FUV} & \colhead{$\sigma_{FUV}$} & \colhead{NUV} &
\colhead{$\sigma_{NUV}$} & \colhead{F255W} & \colhead{$\sigma_{F255W}$}
& \colhead{F300W} & \colhead{$\sigma_{F300W}$} & \colhead{$U$} &
\colhead{$\sigma_{U}$} & \colhead{$B$} & \colhead{$\sigma_{B}$} &
\colhead{$V$} & \colhead{$\sigma_{V}$} & \colhead{$R$} &
\colhead{$\sigma_{R}$} & \colhead{F814W} & \colhead{$\sigma_{F814W}$}
}
\startdata
 001 &       0.354\tablenotemark{2} &   0.022 &   0.208\tablenotemark{2} &   0.006 & \nodata & \nodata &   0.292
  &   0.010 & \nodata & \nodata & \nodata & \nodata & \nodata & \nodata &
 \nodata & \nodata & \nodata & \nodata \\
 002 &     \nodata & \nodata & \nodata & \nodata & \nodata & \nodata &   0.383
  &   0.013 & \nodata & \nodata & \nodata & \nodata & \nodata & \nodata &
 \nodata & \nodata &   0.180 &   0.005 \\
 003 &     \nodata & \nodata & \nodata & \nodata & \nodata & \nodata & \nodata
  & \nodata &   0.373 &   0.049 &   0.368 &   0.025 &   0.391 &   0.018 &
   0.226 &   0.092 & \nodata & \nodata \\
 004 &     \nodata & \nodata & \nodata & \nodata & \nodata & \nodata &   0.745
  &   0.005 & \nodata & \nodata & \nodata & \nodata & \nodata & \nodata &
 \nodata & \nodata &   0.359 &   0.003 \\
 005 &     \nodata & \nodata & \nodata & \nodata & \nodata & \nodata & \nodata
  & \nodata &   0.309\tablenotemark{2} &   0.069 &   0.479\tablenotemark{2} &   0.019 &   0.371\tablenotemark{2} &   0.017 &
   0.245\tablenotemark{2} &   0.043 & \nodata & \nodata \\
 006 &       0.635\tablenotemark{2} &   0.001 &   0.534\tablenotemark{2} &   0.001 & \nodata & \nodata &   0.668
  &   0.005 & \nodata & \nodata & \nodata & \nodata & \nodata & \nodata &
 \nodata & \nodata &   0.293 &   0.006 \\
 007 &     \nodata & \nodata & \nodata & \nodata & \nodata & \nodata & \nodata
  & \nodata &   0.504 &   0.035 &   0.541 &   0.016 &   0.539 &   0.011 &
   0.460 &   0.030 & \nodata & \nodata \\
 008 &     \nodata & \nodata & \nodata & \nodata & \nodata & \nodata & \nodata
  & \nodata &   0.329\tablenotemark{2} &   0.034 &   0.273\tablenotemark{2} &   0.026 &   0.271\tablenotemark{2} &   0.008 &
   0.248\tablenotemark{2} &   0.006 & \nodata & \nodata \\
 009 &     \nodata & \nodata & \nodata & \nodata & \nodata & \nodata &   0.264
  &   0.038 & \nodata & \nodata & \nodata & \nodata & \nodata & \nodata &
 \nodata & \nodata & \nodata & \nodata \\
 010 &     \nodata & \nodata & \nodata & \nodata & \nodata & \nodata & \nodata
  & \nodata &   0.424 &   0.021 &   0.416 &   0.016 &   0.379 &   0.009 &
   0.407 &   0.006 & \nodata & \nodata \\
 011 &     \nodata & \nodata & \nodata & \nodata & \nodata & \nodata &   0.306
  &   0.013 & \nodata & \nodata & \nodata & \nodata & \nodata & \nodata &
 \nodata & \nodata & \nodata & \nodata \\
 012 &     \nodata & \nodata & \nodata & \nodata & \nodata & \nodata & \nodata
  & \nodata &   0.448\tablenotemark{2} &   0.028 &   0.524\tablenotemark{2} &   0.004 &   0.461\tablenotemark{2} &   0.004 &
   0.444\tablenotemark{2} &   0.007 & \nodata & \nodata \\
 013 &     \nodata & \nodata &   0.348\tablenotemark{2} &   0.064 & \nodata & \nodata &   1.480
  &   0.027 & \nodata & \nodata & \nodata & \nodata & \nodata & \nodata &
 \nodata & \nodata &   0.620 &   0.004 \\
 014 &       0.380 &   0.009 &   0.287 &   0.005 & \nodata & \nodata &   0.323
  &   0.041 &   0.235 &   0.024 &   0.150 &   0.036 &   0.157 &   0.021 &
   0.171 &   0.022 &   0.132 &   0.018 \\
 015 &     \nodata & \nodata & \nodata & \nodata & \nodata & \nodata & \nodata
  & \nodata &   0.211 &   0.176 &   0.235 &   0.118 &   0.350 &   0.074 &
   0.073 &   0.129 & \nodata & \nodata \\
 016 &     \nodata & \nodata & \nodata & \nodata & \nodata & \nodata &   0.593
  &   0.004 & \nodata & \nodata & \nodata & \nodata & \nodata & \nodata &
 \nodata & \nodata & \nodata & \nodata \\
 017 &     \nodata & \nodata & \nodata & \nodata & \nodata & \nodata & \nodata
  & \nodata &   0.250 &   0.008 &   0.162 &   0.011 &   0.132 &   0.006 &
   0.119 &   0.004 & \nodata & \nodata \\
 018 &     \nodata & \nodata & \nodata & \nodata & \nodata & \nodata & \nodata
  & \nodata &   0.287 &   0.040 &   0.277 &   0.014 &   0.209 &   0.035 &
   0.158 &   0.025 & \nodata & \nodata \\
 019 &     \nodata & \nodata & \nodata & \nodata & \nodata & \nodata & \nodata
  & \nodata &   0.837 &   0.023 &   0.956 &   0.004 &   0.991 &   0.003 &
   1.034 &   0.004 & \nodata & \nodata \\
 020 &     \nodata & \nodata & \nodata & \nodata & \nodata & \nodata & \nodata
  & \nodata &   0.231 &   0.095 &   0.292 &   0.030 &   0.266 &   0.022 &
   0.211 &   0.041 & \nodata & \nodata \\
 021 &     \nodata & \nodata & \nodata & \nodata & \nodata & \nodata & \nodata
  & \nodata &   0.268 &   0.025 &   0.188 &   0.021 &   0.166 &   0.012 &
   0.156 &   0.010 & \nodata & \nodata \\
 022 &     \nodata & \nodata & \nodata & \nodata & \nodata & \nodata &   0.330
  &   0.000 & \nodata & \nodata & \nodata & \nodata & \nodata & \nodata &
 \nodata & \nodata &   0.132 &   0.001 \\
 023 &     \nodata & \nodata & \nodata & \nodata & \nodata & \nodata &   0.385
  &   0.007 &   0.402 &   0.035 &   0.409 &   0.008 &   0.392 &   0.007 &
   0.373 &   0.005 &   0.219 &   0.003 \\
 024 &     \nodata & \nodata & \nodata & \nodata & \nodata & \nodata &   0.341
  &   0.022 & \nodata & \nodata & \nodata & \nodata & \nodata & \nodata &
 \nodata & \nodata &   0.206 &   0.008 \\
 025 &     \nodata & \nodata & \nodata & \nodata & \nodata & \nodata &   0.605
  &   0.008 & \nodata & \nodata & \nodata & \nodata & \nodata & \nodata &
 \nodata & \nodata &   0.198 &   0.002 \\
\enddata
\tablecomments{{\bf Columns:} Asymmetry indices in each filter
and their uncertainties ($\sigma_{filter}$). The ID\# is the
identification number assigned to each galaxy in Table 1.
The full table is available only in the electronic edition.
}
\tablenotetext{1}{Galaxy does not meet the S/N requirements for
accurate CAS parameter measurements in this image (S/N $< 75$).}
\tablenotetext{2}{Galaxy does not meet the resolution requirements for
accurate CAS parameter measurements in this image (R $> 1.25$ kpc).}
\end{deluxetable*}

%
\begin{deluxetable*}{lrrrrrrrrrrrrrrrrrr}
\tabletypesize{\scriptsize}
\tablecolumns{19}
\tablewidth{\textwidth}
\tablecaption{Clumpiness Index (S)}
\tablehead{
\colhead{ID\#} & \colhead{FUV} & \colhead{$\sigma_{FUV}$} & \colhead{NUV} &
\colhead{$\sigma_{NUV}$} & \colhead{F255W} & \colhead{$\sigma_{F255W}$}
& \colhead{F300W} & \colhead{$\sigma_{F300W}$} & \colhead{$U$} &
\colhead{$\sigma_{U}$} & \colhead{$B$} & \colhead{$\sigma_{B}$} &
\colhead{$V$} & \colhead{$\sigma_{V}$} & \colhead{$R$} &
\colhead{$\sigma_{R}$} & \colhead{F814W} & \colhead{$\sigma_{F814W}$}
}
\startdata
 001 &       0.00\tablenotemark{2} &    0.000 &   0.00\tablenotemark{2} &    0.000 & \nodata & \nodata &  -0.01
  &    0.038 & \nodata & \nodata & \nodata & \nodata & \nodata & \nodata &
 \nodata & \nodata & \nodata & \nodata \\
 002 &     \nodata & \nodata & \nodata & \nodata & \nodata & \nodata &   0.16
  &    0.194 & \nodata & \nodata & \nodata & \nodata & \nodata & \nodata &
 \nodata & \nodata &   0.25 &    0.187 \\
 003 &     \nodata & \nodata & \nodata & \nodata & \nodata & \nodata & \nodata
  & \nodata &   0.75 &    0.092 &   0.44 &    0.078 &   0.53 &    0.057 &
   0.25 &    0.036 & \nodata & \nodata \\
 004 &     \nodata & \nodata & \nodata & \nodata & \nodata & \nodata &   0.57
  &    0.256 & \nodata & \nodata & \nodata & \nodata & \nodata & \nodata &
 \nodata & \nodata &   0.27 &    0.154 \\
 005 &     \nodata & \nodata & \nodata & \nodata & \nodata & \nodata & \nodata
  & \nodata &   0.30\tablenotemark{2} &    0.041 &   0.52\tablenotemark{2} &    0.062 &   0.51\tablenotemark{2} &    0.051 &
   0.22\tablenotemark{2} &    0.026 & \nodata & \nodata \\
 006 &       0.53\tablenotemark{2} &    0.883 &   0.32\tablenotemark{2} &    0.380 & \nodata & \nodata &   0.63
  &    0.225 & \nodata & \nodata & \nodata & \nodata & \nodata & \nodata &
 \nodata & \nodata &   0.37 &    0.162 \\
 007 &     \nodata & \nodata & \nodata & \nodata & \nodata & \nodata & \nodata
  & \nodata &   0.42 &    0.034 &   0.43 &    0.035 &   0.42 &    0.032 &
   0.18 &    0.015 & \nodata & \nodata \\
 008 &     \nodata & \nodata & \nodata & \nodata & \nodata & \nodata & \nodata
  & \nodata &   0.50\tablenotemark{2} &    0.037 &   0.32\tablenotemark{2} &    0.030 &   0.32\tablenotemark{2} &    0.021 &
   0.34\tablenotemark{2} &    0.017 & \nodata & \nodata \\
 009 &     \nodata & \nodata & \nodata & \nodata & \nodata & \nodata &  -0.40
  &    0.468 & \nodata & \nodata & \nodata & \nodata & \nodata & \nodata &
 \nodata & \nodata & \nodata & \nodata \\
 010 &     \nodata & \nodata & \nodata & \nodata & \nodata & \nodata & \nodata
  & \nodata &   0.67 &    0.019 &   0.61 &    0.028 &   0.53 &    0.019 &
   0.64 &    0.018 & \nodata & \nodata \\
 011 &     \nodata & \nodata & \nodata & \nodata & \nodata & \nodata &   0.50
  &    0.332 & \nodata & \nodata & \nodata & \nodata & \nodata & \nodata &
 \nodata & \nodata & \nodata & \nodata \\
 012 &     \nodata & \nodata & \nodata & \nodata & \nodata & \nodata & \nodata
  & \nodata &   0.27\tablenotemark{2} &    0.023 &   0.40\tablenotemark{2} &    0.031 &   0.35\tablenotemark{2} &    0.022 &
   0.38\tablenotemark{2} &    0.017 & \nodata & \nodata \\
 013 &     \nodata & \nodata &   0.57\tablenotemark{2} &    2.095 & \nodata & \nodata &   1.01
  &    0.628 & \nodata & \nodata & \nodata & \nodata & \nodata & \nodata &
 \nodata & \nodata &   0.32 &    0.156 \\
 014 &       0.81 &    1.873 &   0.28 &    0.619 & \nodata & \nodata &   0.61
  &    0.448 &   0.20 &    0.017 &   0.15 &    0.018 &   0.17 &    0.016 &
   0.11 &    0.013 &  -0.10 &    0.144 \\
 015 &     \nodata & \nodata & \nodata & \nodata & \nodata & \nodata & \nodata
  & \nodata &   0.18 &    0.048 &   0.38 &    0.090 &   0.44 &    0.082 &
  -0.84 &    0.094 & \nodata & \nodata \\
 016 &     \nodata & \nodata & \nodata & \nodata & \nodata & \nodata &   0.71
  &    0.188 & \nodata & \nodata & \nodata & \nodata & \nodata & \nodata &
 \nodata & \nodata & \nodata & \nodata \\
 017 &     \nodata & \nodata & \nodata & \nodata & \nodata & \nodata & \nodata
  & \nodata &   0.69 &    0.030 &   0.42 &    0.019 &   0.32 &    0.010 &
   0.31 &    0.007 & \nodata & \nodata \\
 018 &     \nodata & \nodata & \nodata & \nodata & \nodata & \nodata & \nodata
  & \nodata &   0.58 &    0.038 &   0.38 &    0.034 &   0.31 &    0.025 &
   0.08 &    0.010 & \nodata & \nodata \\
 019 &     \nodata & \nodata & \nodata & \nodata & \nodata & \nodata & \nodata
  & \nodata &   0.81 &    0.024 &   0.99 &    0.030 &   0.97 &    0.023 &
   0.93 &    0.018 & \nodata & \nodata \\
 020 &     \nodata & \nodata & \nodata & \nodata & \nodata & \nodata & \nodata
  & \nodata &   0.36 &    0.037 &   0.40 &    0.045 &   0.38 &    0.036 &
   0.31 &    0.027 & \nodata & \nodata \\
 021 &     \nodata & \nodata & \nodata & \nodata & \nodata & \nodata & \nodata
  & \nodata &   0.68 &    0.039 &   0.40 &    0.034 &   0.33 &    0.024 &
   0.31 &    0.019 & \nodata & \nodata \\
 022 &     \nodata & \nodata & \nodata & \nodata & \nodata & \nodata &   0.30
  &    0.035 & \nodata & \nodata & \nodata & \nodata & \nodata & \nodata &
 \nodata & \nodata &   0.35 &    0.056 \\
 023 &     \nodata & \nodata & \nodata & \nodata & \nodata & \nodata &   0.04
  &    0.076 &   0.75 &    0.035 &   0.65 &    0.038 &   0.59 &    0.028 &
   0.56 &    0.023 &   0.16 &    0.132 \\
 024 &     \nodata & \nodata & \nodata & \nodata & \nodata & \nodata &   0.51
  &    0.353 & \nodata & \nodata & \nodata & \nodata & \nodata & \nodata &
 \nodata & \nodata &   0.15 &    0.162 \\
 025 &     \nodata & \nodata & \nodata & \nodata & \nodata & \nodata &   0.77
  &    0.265 & \nodata & \nodata & \nodata & \nodata & \nodata & \nodata &
 \nodata & \nodata &   0.20 &    0.104 \\
\enddata
\tablecomments{{\bf Columns:} Clumpiness indices in each filter
and their uncertainties ($\sigma_{filter}$). The ID\# is the
identification number assigned to each galaxy in Table 1.
The full table is available only in the electronic edition.
}
\tablenotetext{1}{Galaxy does not meet the S/N requirements for
accurate CAS parameter measurements in this image (S/N $< 75$).}
\tablenotetext{2}{Galaxy does not meet the resolution requirements for
accurate CAS parameter measurements in this image (R $> 1.25$ kpc).}
\end{deluxetable*}

\begin{figure*}
\plotone{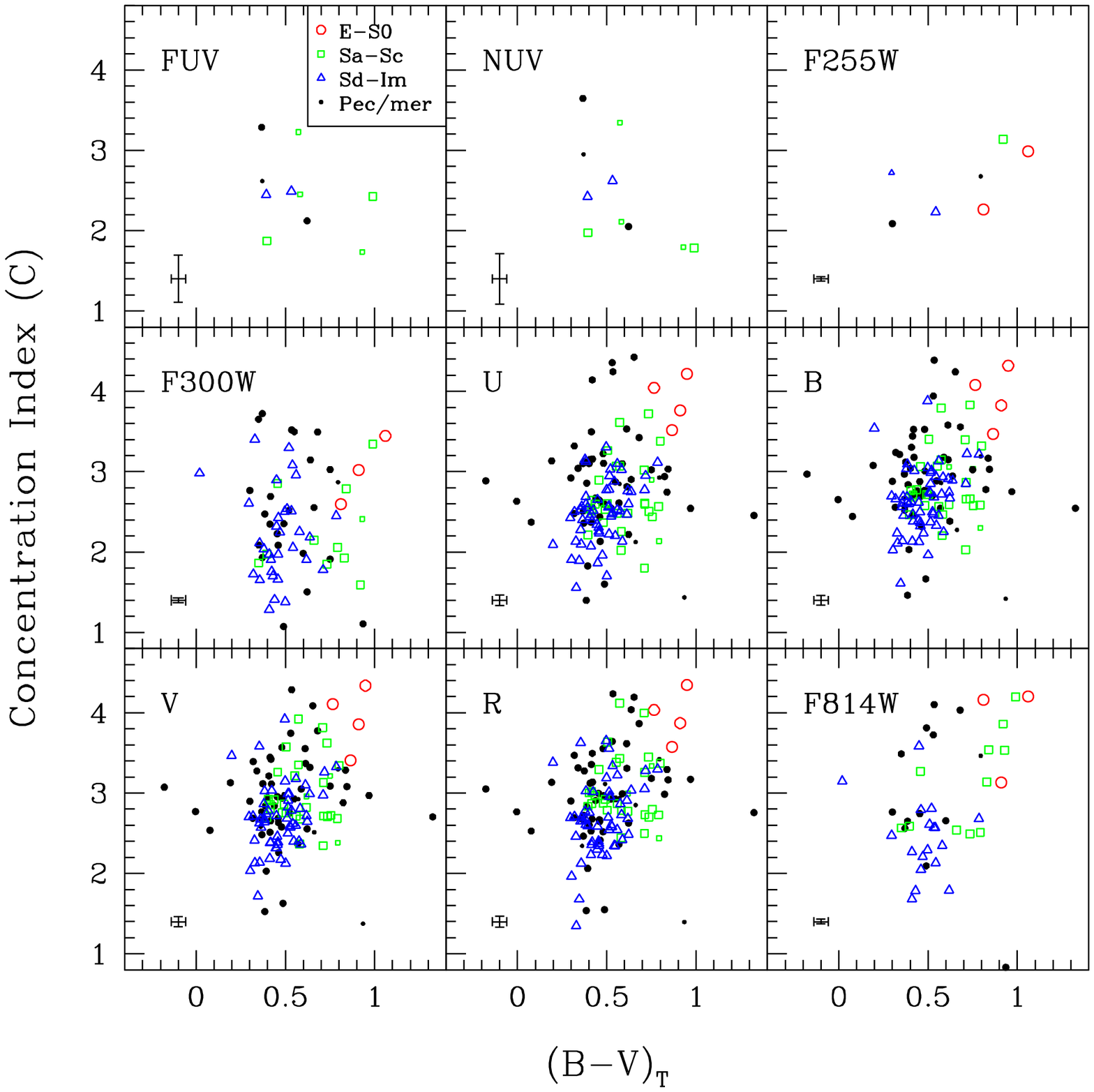}
\figcaption[f4.eps]{{Concentration index as a function
of total ($B-V$) color for nine different pass-bands.
The symbols are coded by galaxy type, as indicated in the upper left
(FUV) panel. Smaller symbols represent galaxies that do not meet our
adopted reliability criteria (see text). Vertical error
bars in the lower left corner of each panel represent the median error
on the individual concentration index values. Horizontal error bars
show the median error on ($B-V$). In general, redder galaxies
tend to be more concentrated. This trend also results in a
segregation according to galaxy type.
}}
\end{figure*}

\begin{figure*}
\plotone{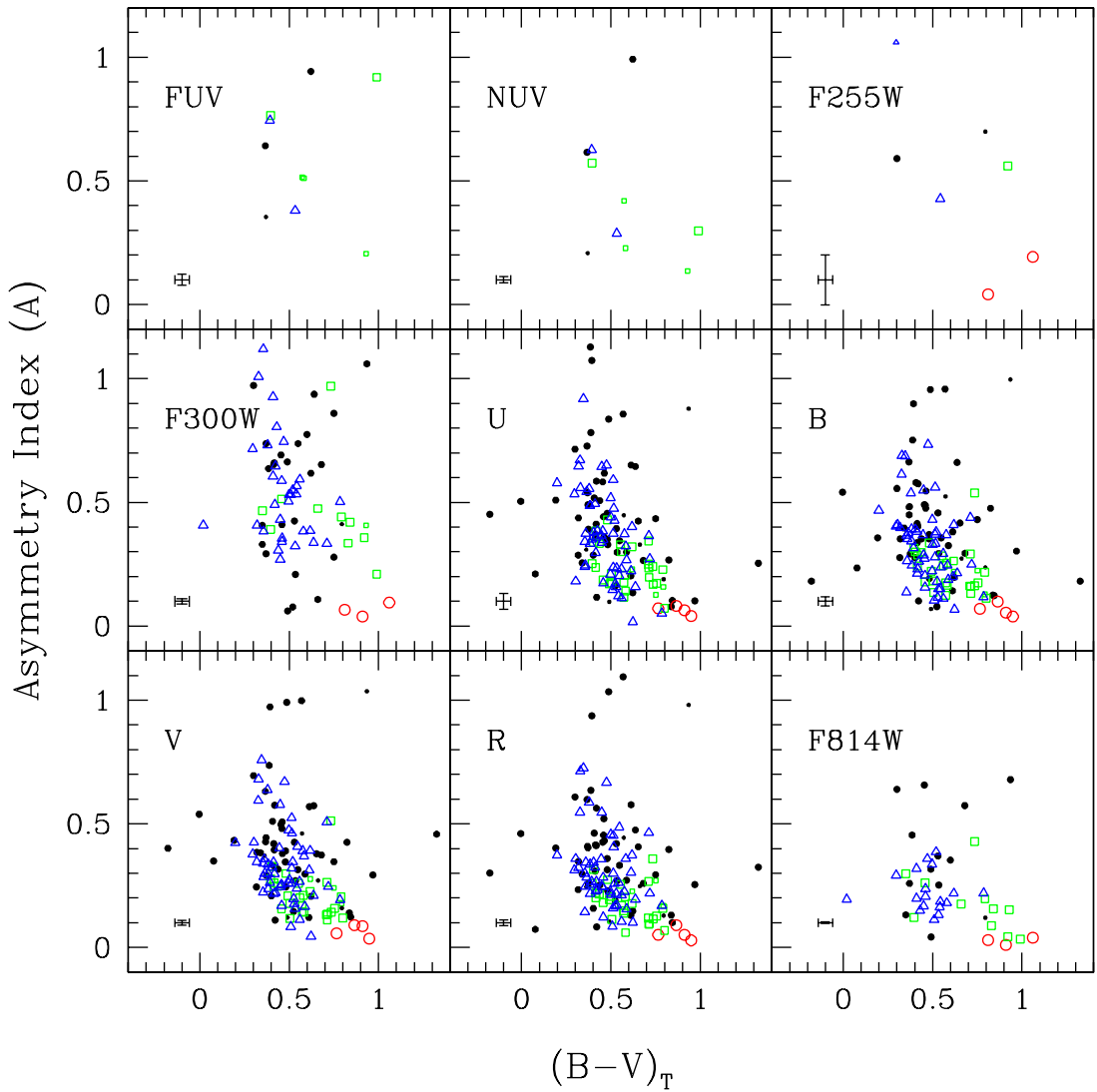}
\figcaption[f5.eps]{{Asymmetry index as a function of total
($B-V$) color and filter. The symbols are coded as in Fig.~4. Error
bars in the lower left corner of each panel represent the median error on
the individual asymmetry values and on ($B-V$). In general, redder
galaxies tend to
be more symmetric. As in Fig.~4, this trend with galaxy color also results
in a segregation according to galaxy type. Peculiar/merging galaxies are
in general slightly bluer and more asymmetric than normal galaxies.
}}
\end{figure*}

\begin{figure*}
\plotone{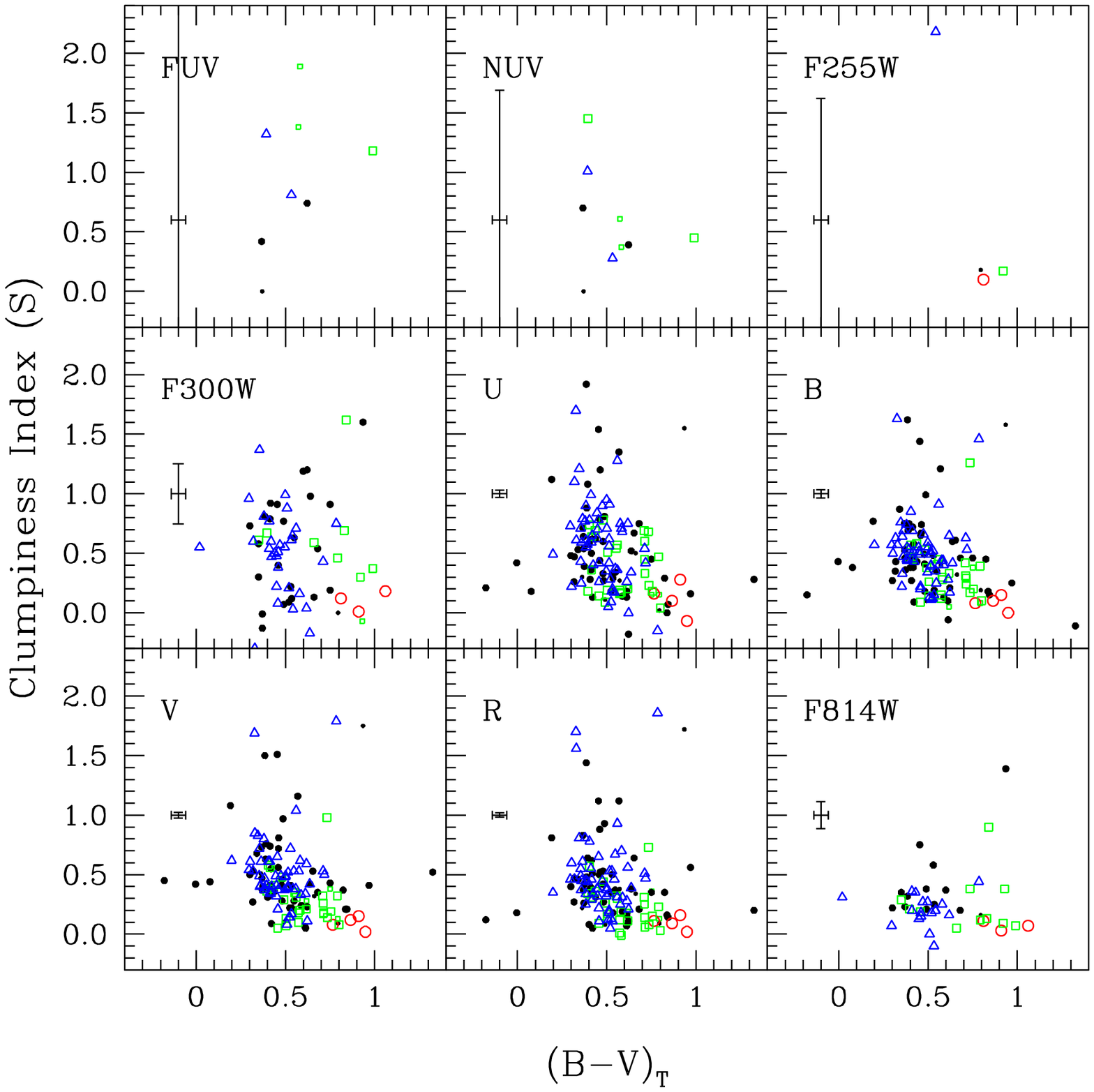}
\figcaption[f6.eps]{{Clumpiness index as a function of
total ($B-V$) color and filter. The symbols are coded as in Fig.~4.
Error bars on the left side of each panel represent the median
error on the individual clumpiness values, and on ($B-V$). The large
errors in S for FUV and NUV are due to
the low resolution of the GALEX images (stellar FWHM $\gtrsim 5\arcsec$),
while the large errors in F255W and F300W are due to poor
S/N and CTE effects. In general, redder galaxies tend to be
less clumpy. As in Figs. 4 and 5, this trend with galaxy color also
results in a segregation according to galaxy type.
There is a larger spread in S at shorter wavelengths, with blue galaxies
appearing more clumpy at shorter wavelengths than they do at longer
wavelengths. Peculiar/merging galaxies are in general slightly
bluer and more clumpy than normal galaxies.
}}
\end{figure*}

Figures 7, 8, and 9 show the C, A, and S parameters, respectively, of
each galaxy as a function of galaxy type. As for Figures 4--6, 
separate panels show results from
different filters. Data-points from galaxies that did not meet the reliability
criteria (S/N$<75$, or R$>1.25$ kpc) are indicated with small black
points. Median values of the CAS parameters in a particular
filter within a particular type-bin are indicated with large colored symbols,
and the spread of the data points is indicated by the 25--75\% quartile
range, represented by the vertical error bars. These
median C, A, and S values are also listed in Tables 6, 7, and 8,
respectively. Data that did not meet the reliability criteria 
were not included in the 
medians. Due to low S/N and significant CTE effects (particularly in F255W) 
causing the clumpiness measured in the sky to be higher than that measured
in some galaxies, median values of S were only calculated with data that had 
S$>-0.2$. In all three plots, the spread in the CAS parameters
increases toward later galaxy type.  
There are few objects in the GALEX FUV and NUV filters, as well as the
HST F255W filter, but the other filters show a trend of decreasing
C, and increasing A and S toward later type for normal galaxies 
(E through Im).
The slope of the dependence of A and S on type for normal galaxies decreases 
toward longer wavelengths, such that late-type galaxies are more 
asymmetric and clumpy at shorter wavelengths than at longer wavelengths,
which is also seen in Figures 4--6. The peculiar/merging galaxies are 
plotted as their sub-types, as defined in Section 4.1, in Figures 7--9.
Merging galaxies (solid triangles) tend to be in general much less 
concentrated and much more asymmetric and clumpy than any other 
galaxy type. Pre-merging
galaxies (solid circles) have similar or slightly lower C and higher
A and S values than normal mid- to late-type galaxies, because these galaxies
are currently only slightly distorted by the tidal interactions with their 
neighbors. Minor
mergers (solid squares) are slightly more concentrated and less asymmetric 
and clumpy than the pre-mergers, because the low mass of the smaller 
interacting galaxy has likely less of an effect on the brighter, 
larger galaxy's light distribution 
than a neighboring galaxy of similar size would have had. This is also
found in minor vs. major merger simulations (Conselice 2006).
 Merger remnants 
(asterisks) are more concentrated, more asymmetric, and similar in 
clumpiness to pre-mergers and normal mid- to late-type galaxies. 
Peculiar galaxies (crosses) have similar CAS values as normal 
mid- to late-type galaxies, although they appear slightly more 
concentrated and more asymmetric in some filters.

%
\begin{deluxetable*}{llllllllll}
\tabletypesize{\scriptsize}
\tablecolumns{10}
\tablewidth{5in}
\tablecaption{Median C For Each Galaxy Type Bin}
\tablehead{
\colhead{Type} & \colhead{FUV} & \colhead{NUV}
& \colhead{F255W} &
\colhead{F300W} & \colhead{$U$} &
\colhead{$B$} & \colhead{$V$} & \colhead{$R$} &
\colhead{F814W}
}
\startdata
 E--S0 &  \nodata & 1.73(1) & 2.63(2) & 2.93(4) & 3.90(4) & 3.95(4) & 3.98(4)
  & 3.95(4) & 3.65(4) \\
   &  \nodata & \nodata & 0.36 & 0.26 & 0.24 & 0.28 & 0.30 & 0.23 & 0.55
  \\
 Sa--Sc &  2.15(2) & 1.88(2) & 3.14(1) & 2.06(11) & 2.58(26) & 2.71(26) &
 2.84(26) & 2.90(26) & 3.18(12) \\
   &  0.28 & 0.09 & \nodata & 0.46 & 0.25 & 0.24 & 0.26 & 0.28 & 0.49
  \\
 Sd--Im &  2.47(2) & 2.52(2) & 2.23(1) & 2.11(41) & 2.50(55) & 2.63(55) &
 2.67(55) & 2.65(55) & 2.35(25) \\
   &  0.02 & 0.10 & \nodata & 0.35 & 0.26 & 0.27 & 0.29 & 0.32 & 0.28
  \\
 pM &  2.12(1) & 2.05(1) & \nodata & 2.22(8) & 2.60(7) & 2.77(7) & 2.58(7) &
 2.63(7) & 2.86(4) \\
  &  \nodata & \nodata & \nodata & 0.35 & 0.36 & 0.36 & 0.39 & 0.39 &
 0.53 \\
 mM &  \nodata & \nodata & 1.89(1) & 1.94(3) & 3.03(3) & 3.03(3) & 3.08(3) &
 3.17(3) & 2.56(3) \\
   &  \nodata & \nodata & \nodata & 0.79 & 0.94 & 0.91 & 0.81 & 0.76 &
 1.25 \\
 M &  \nodata & \nodata & 2.30(2) & 1.44(6) & 2.61(3) & 2.54(3) & 2.37(3) &
 2.36(3) & 1.47(4) \\
   &  \nodata & \nodata & 0.15 & 0.67 & 0.63 & 0.59 & 0.61 & 0.64 & 0.79
  \\
 MR &  3.29(1) & 3.65(1) & \nodata & 2.69(5) & 3.04(15) & 3.15(15) & 3.21(15)
  & 3.18(15) & 3.19(2) \\
   &  \nodata & \nodata & \nodata & 0.34 & 0.20 & 0.26 & 0.31 & 0.44 &
 0.54 \\
 P &  \nodata & \nodata & 2.09(1) & 2.90(8) & 2.67(21) & 2.88(21) & 2.93(21) &
 2.93(21) & 3.13(4) \\
   &  \nodata & \nodata & \nodata & 0.59 & 0.28 & 0.17 & 0.18 & 0.22 &
 0.50 \\
\enddata
\tablecomments{Median concentration indices in each type-bin and filter.
Galaxies that did not meet the reliability criteria (S/N$<75$, or R$>1.25$ kpc)
were not included in the median. Values in parenthesis give the number
of galaxies used for that type-bin and filter. The second line for each
type-bin lists
half the range of the data within the 25 and 75\% quartiles. When
only one data point was available, the value listed is for that data
point and the associated quartile range cannot be computed. The
peculiar/merging galaxies are broken up into sub-types as follows:
pre-merger (pM), minor merger (mM), merger (M), merger remnant (MR), and
peculiar (P). In general, C increases with increasing rest-frame
wavelength.}
\end{deluxetable*}

%

\begin{deluxetable*}{llllllllll}
\tabletypesize{\scriptsize}
\tablecolumns{10}
\tablewidth{5in}
\tablecaption{Median A For Each Galaxy Type Bin}
\tablehead{
\colhead{Type} & \colhead{FUV} & \colhead{NUV}
& \colhead{F255W} &
\colhead{F300W} & \colhead{$U$} &
\colhead{$B$} & \colhead{$V$} & \colhead{$R$} &
\colhead{F814W}
}
\startdata
 E--S0 &  \nodata & 0.19(1) & 0.12(2) & 0.08(4) & 0.07(4) & 0.06(4) & 0.07(4)
  & 0.05(4) & 0.03(4) \\
  &  \nodata & \nodata & 0.08 & 0.04 & 0.01 & 0.02 & 0.02 & 0.02 & 0.01
  \\
 Sa--Sc &  0.84(2) & 0.44(2) & 0.56(1) & 0.42(11) & 0.23(26) & 0.22(26) &
 0.20(26) & 0.18(26) & 0.15(12) \\
   &  0.08 & 0.14 & \nodata & 0.07 & 0.07 & 0.06 & 0.06 & 0.05 & 0.06
  \\
 Sd--Im &  0.56(2) & 0.46(2) & 0.43(1) & 0.50(41) & 0.36(55) & 0.32(55) &
 0.30(55) & 0.27(55) & 0.22(25) \\
   &  0.18 & 0.17 & \nodata & 0.11 & 0.13 & 0.09 & 0.08 & 0.07 & 0.06
  \\
 pM &  0.94(1) & 0.99(1) & \nodata & 0.64(8) & 0.36(7) & 0.34(7) & 0.33(7) &
 0.31(7) & 0.34(4) \\
   &  \nodata & \nodata & \nodata & 0.24 & 0.16 & 0.11 & 0.11 & 0.13 &
 0.15 \\
 mM &  \nodata & \nodata & 0.78(1) & 0.74(3) & 0.30(3) & 0.29(3) & 0.27(3) &
 0.27(3) & 0.26(3) \\
   &  \nodata & \nodata & \nodata & 0.53 & 0.19 & 0.18 & 0.16 & 0.15 &
 0.43 \\
 M &  \nodata & \nodata & 0.80(2) & 1.09(6) & 0.84(3) & 0.96(3) & 0.99(3) &
 1.03(3) & 0.84(4) \\
   &  \nodata & \nodata & 0.19 & 0.14 & 0.04 & 0.10 & 0.13 & 0.23 & 0.07
  \\
 MR &  0.64(1) & 0.62(1) & \nodata & 0.66(5) & 0.43(15) & 0.43(15) & 0.43(15)
  & 0.41(15) & 0.41(2) \\
   &  \nodata & \nodata & \nodata & 0.11 & 0.15 & 0.14 & 0.10 & 0.13 &
 0.04 \\
 P &  \nodata & \nodata & 0.59(1) & 0.37(8) & 0.36(21) & 0.30(21) & 0.37(21) &
 0.30(21) & 0.61(4) \\
   &  \nodata & \nodata & \nodata & 0.19 & 0.12 & 0.13 & 0.10 & 0.10 &
 0.15 \\
\enddata
\tablecomments{Median asymmetry indices in each type-bin and filter.
Galaxies that did not meet the reliability criteria (S/N$<75$, or R$>1.25$ kpc)
were not included in the median. Values in parenthesis give the number
of galaxies used for that type-bin and filter. The second line for each
type-bin lists
half the range of the data within the 25 and 75\% quartiles. Median
values that were only calculated from one data point do not have an
associated quartile range listed. The
peculiar/merging galaxies are broken up into sub-types as follows:
pre-merger (pM), minor merger (mM), merger (M), merger remnant (MR), and
peculiar (P). In general, A decreases with increasing rest-frame
wavelength.}
\end{deluxetable*}

%

\begin{deluxetable*}{llllllllll}
\tabletypesize{\scriptsize}
\tablecolumns{10}
\tablewidth{5in}
\tablecaption{Median S For Each Galaxy Type Bin}
\tablehead{
\colhead{Type} & \colhead{FUV} & \colhead{NUV}
& \colhead{F255W} &
\colhead{F300W} & \colhead{$U$} &
\colhead{$B$} & \colhead{$V$} & \colhead{$R$} &
\colhead{F814W}
}
\startdata
 E--S0 &  \nodata &  2.12(1) & 0.10(1) & 0.12(3) & 0.13(4) & 0.09(4) & 0.10(4)
  & 0.10(4) & 0.05(4) \\
  &  \nodata & \nodata & \nodata & 0.09 & 0.10 & 0.04 & 0.04 & 0.04 &
 0.03 \\
 Sa--Sc &  1.99(2) & 0.95(2) & 0.17(1) & 0.60(10) & 0.35(26) & 0.32(26) &
 0.24(26) & 0.21(26) & 0.16(12) \\
  &  0.81 & 0.50 & \nodata & 0.17 & 0.21 & 0.13 & 0.10 & 0.11 & 0.12
  \\
 Sd--Im &  1.06(2) & 0.65(2) &  \nodata & 0.55(37) & 0.59(55) & 0.50(55) &
 0.48(55) & 0.44(55) & 0.20(25) \\
  &  0.26 & 0.37 & \nodata & 0.16 & 0.21 & 0.10 & 0.12 & 0.12 & 0.06
  \\
 pM &  0.74(1) & 0.39(1) & \nodata & 0.78(8) & 0.36(7) & 0.45(7) & 0.41(7) &
 0.37(7) & 0.35(4) \\
  &  \nodata & \nodata & \nodata & 0.38 & 0.30 & 0.18 & 0.12 & 0.18 &
 0.06 \\
 mM &  \nodata & \nodata & \nodata & 0.12(3) & 0.29(3) & 0.19(3) & 0.22(3) &
 0.18(3) & 0.25(3) \\
  &  \nodata & \nodata & \nodata & 0.21 & 0.24 & 0.15 & 0.08 & 0.06 &
 0.11 \\
 M &  \nodata & \nodata & 0.94(2) & 1.22(6) & 0.88(3) & 0.99(3) & 0.97(3) &
 0.93(3) & 0.75(4) \\
  &  \nodata & \nodata & 0.25 & 0.32 & 0.27 & 0.26 & 0.27 & 0.32 & 0.30
  \\
 MR &  0.42(1) & 0.70(1) & \nodata & 0.91(5) & 0.45(15) & 0.46(15) & 0.43(15)
  & 0.42(15) & 0.45(2) \\
  &  \nodata & \nodata & \nodata & 0.06 & 0.19 & 0.18 & 0.17 & 0.15 &
 0.13 \\
 P &  \nodata & \nodata & \nodata & 0.35(8) & 0.42(21) & 0.38(21) & 0.42(21) &
 0.36(20) & 0.29(4) \\
  &  \nodata & \nodata & \nodata & 0.24 & 0.18 & 0.17 & 0.12 & 0.15 &
 0.17 \\
\enddata
\tablecomments{Median clumpiness indices in each type-bin and filter.
Galaxies that did not meet the reliability criteria (S/N$<75$, or R$>1.25$ kpc)
were not included in the median, as well as values of S$<-0.2$.
Values in parenthesis give the number of galaxies used for that type-bin
and filter. The second line for each type-bin lists
half the range of the data within the 25 and 75\% quartiles. Median
values that were only calculated from one data point do not have an
associated quartile range listed. The
peculiar/merging galaxies are broken up into sub-types as follows:
pre-merger (pM), minor merger (mM), merger (M), merger remnant (MR), and
peculiar (P). In general, S decreases with increasing rest-frame
wavelength.}
\end{deluxetable*}

\begin{figure*}
\plotone{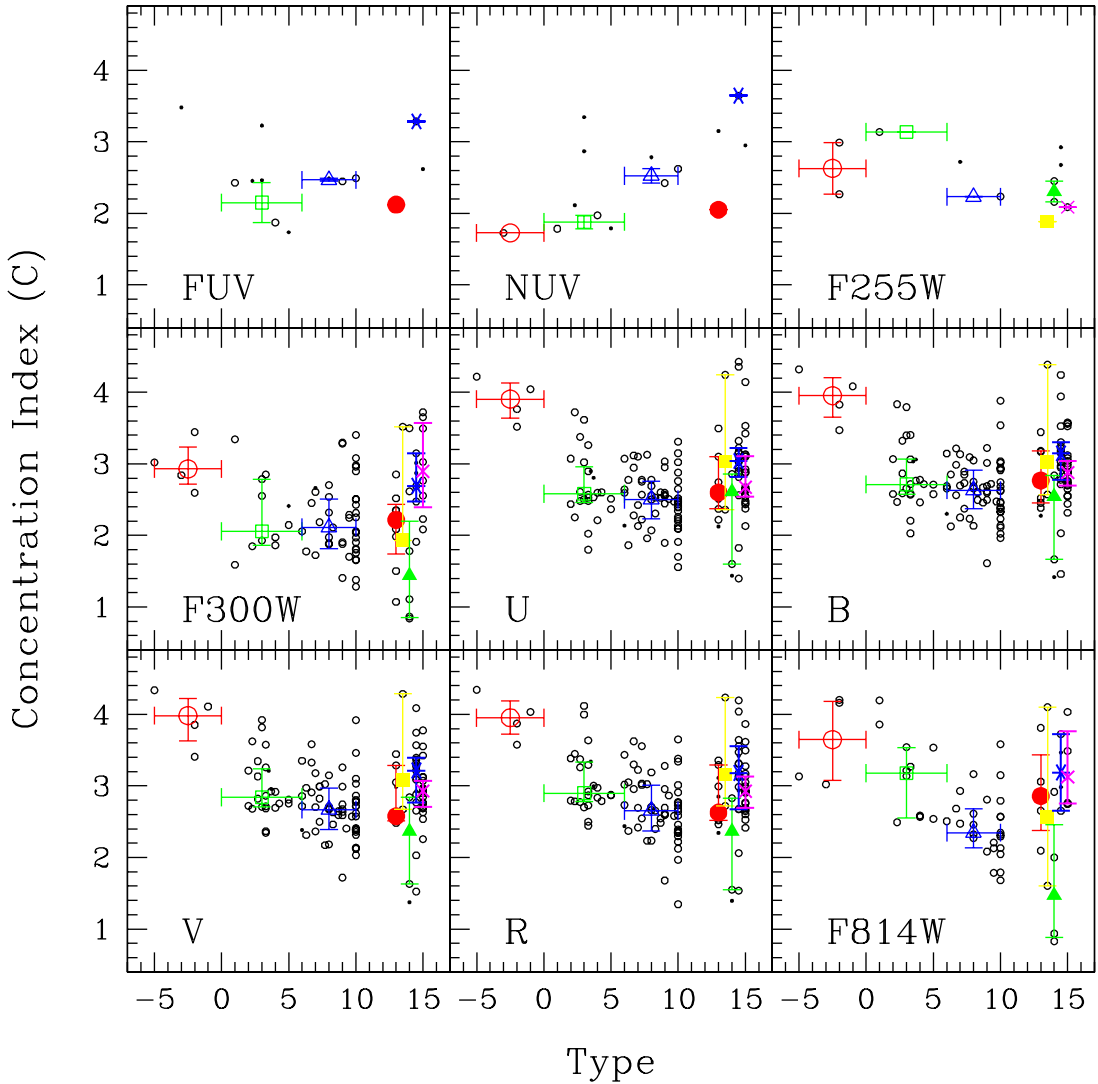}
\figcaption[f7.eps]{{Concentration index as a function
of galaxy type and filter. Small points represent
galaxies that do not meet our adopted reliability criteria (see text).
The large colored symbols represent the median C values for a
type-bin whose width is defined by the horizontal error
bars. The vertical error bars show the 25--75\% quartile ranges. These
same symbols are used in subsequent figures to represent these galaxy
types (as labeled in Fig.~15). There is a larger spread in C with later
galaxy type, with galaxies tending on average to be less concentrated
with later normal galaxy type (E through Im). Merging galaxies (T = 14)
are in general less concentrated than all other galaxy types.
}}
\end{figure*}

\begin{figure*}
\plotone{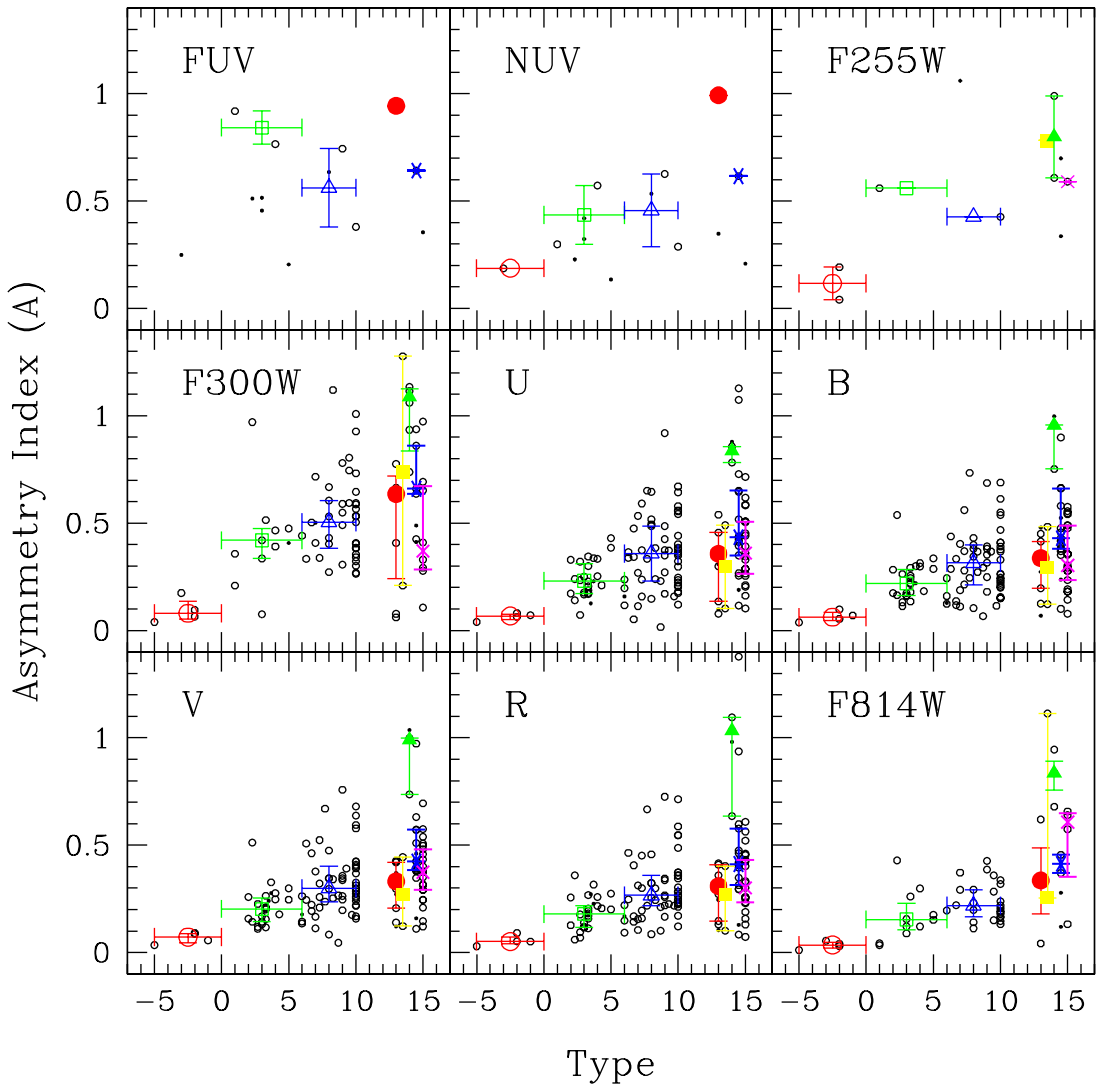}
\figcaption[f8.eps]{{Asymmetry index as a function of
galaxy type. Symbols are coded as in Fig.~7.
The large colored symbols represent the median A values for each galaxy type
bin, with error bar as in Fig.~7. There is a
larger spread in A for later galaxy types, with galaxies tending
on average to be less symmetric with later normal galaxy type (E through Im).
Late-type galaxies are less symmetric at shorter wavelengths than at
longer wavelengths. Merging galaxies (T = 14) are in general less symmetric
than all other galaxy types.
}}
\end{figure*}

\begin{figure*}
\plotone{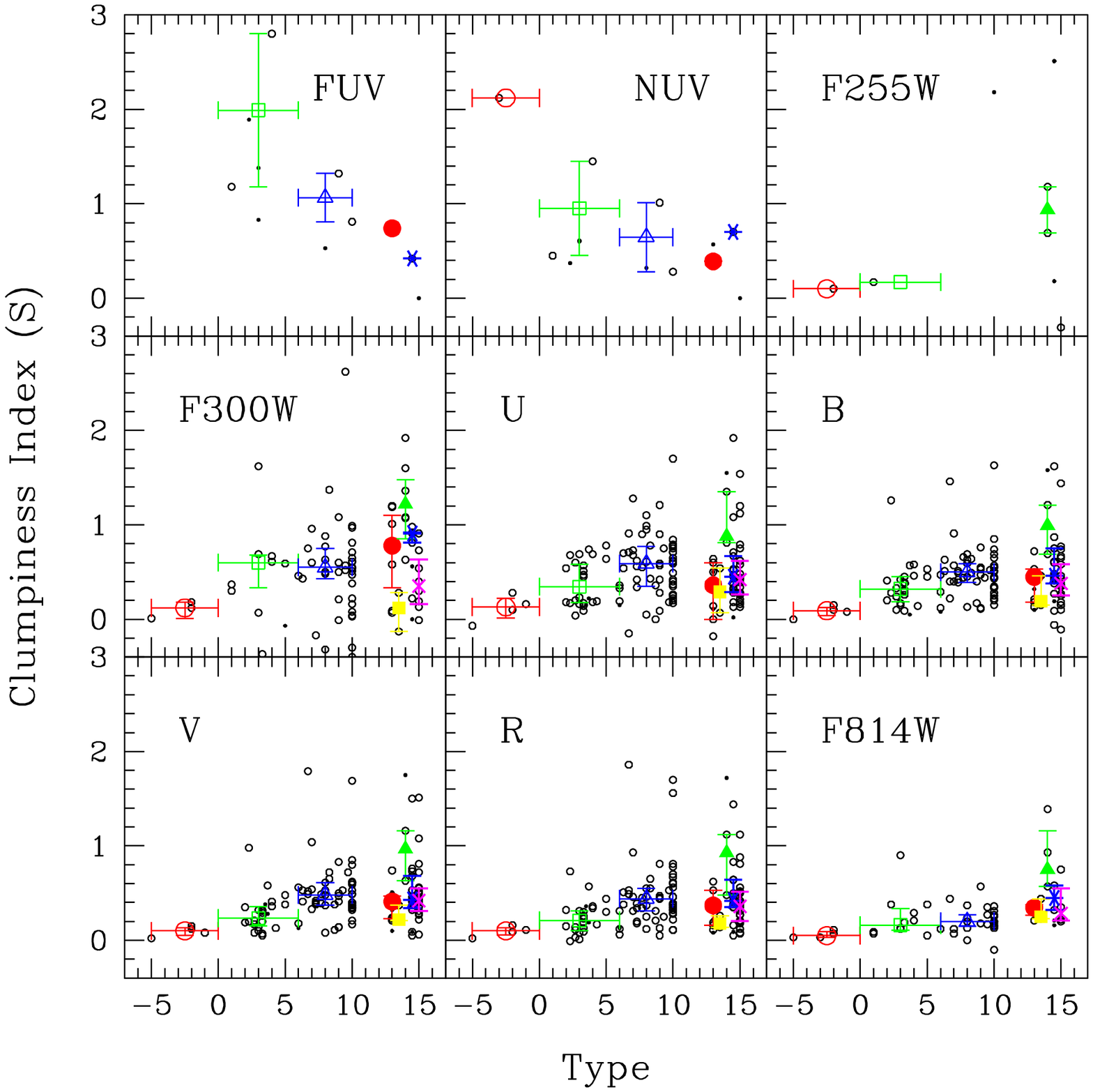}
\figcaption[f9.eps]{{Clumpiness index as a function of
galaxy type. Symbols are coded as in Fig.~7.
The large colored symbols represent the median S values for each type
bin, with error bars as in Fig.~7. An additional
criteria that S$>-0.2$ was used in the calculation of the median value.
There is a
larger spread in S with later galaxy type, with galaxies tending
on average to be more clumpy with later normal galaxy type (E through Im).
Late-type galaxies are more clumpy at shorter wavelengths than at
longer wavelengths. Merging galaxies (T = 14) are in general more
clumpy than any other galaxy type.
}}
\end{figure*}

Histograms of the distribution of the CAS parameters as separated by
galaxy type are presented in Figure 10. These distributions demonstrate
the amount of overlap in the CAS parameters between different galaxy
type-bins, and therefore are relevant to help judge the reliability of 
using these parameters to classify galaxies. The left panels of this 
plot contain all data in filters with wavelengths shortward of the Balmer 
break ($\lambda_c \lesssim 360$~nm), and the right panels contain all data in 
filters with wavelengths longward of the Balmer break ($\lambda_c \gtrsim 
360$~nm).
Galaxies that did not meet our reliability criteria were not included
in this plot. There is a large amount of overlap between the concentration
indices of all galaxy types, particularly at shorter wavelengths. Early
galaxy types are better separated from other galaxy types by 
concentration at longer wavelengths, although there is still a
significant amount of overlap for smaller values of C. Very few galaxies
with types later than S0 have high values of C. Therefore, there is
a high probability that a galaxy with high concentration index 
(C $\gtrsim 3.5$) is
an early-type (E--S0) galaxy. Similarly, early-type (E--S0) galaxies
have very little overlap with later galaxy types in asymmetry index,
particularly at longer wavelengths. Galaxies with A $\lesssim 0.1$ are likely
early-types (E--S0), which is essentially true at all wavelengths.
Although all early-type (E--S0) galaxies
have low clumpiness indices, there is a larger overlap with later-type
galaxies for this index than for the concentration and asymmetry indices.
In general, almost all early-type (E--S0) galaxies have S $\lesssim 0.2$,
but not all galaxies with S $\lesssim 0.2$ are early-types. There is a single
early-type galaxy with a high S measurement in the NUV, but this is
likely due to low S/N in the parts of the galaxy just outside its bright
inner core.
Although all of the later (Sa--Im and peculiar/merging) galaxy type 
distributions are offset slightly from each other in mean C, A, and S, there
is a large overlap between their CAS parameter values. Therefore,
the CAS parameters cannot be used independently to classify individual
galaxies with types later than Sa, but they can be used to describe something
about the overall distribution of types as a whole in a large sample.
Nonetheless, it is note-worthy that a subset of the peculiar/merging galaxies,
predominantly major mergers, display high asymmetry indices at longer
wavelengths that are not seen for the other galaxy types.
The next two sections discuss using a combination of the CAS parameters as
a more reliable way of determining the morphological distributions within
a galaxy sample.

\begin{figure*}
\plotone{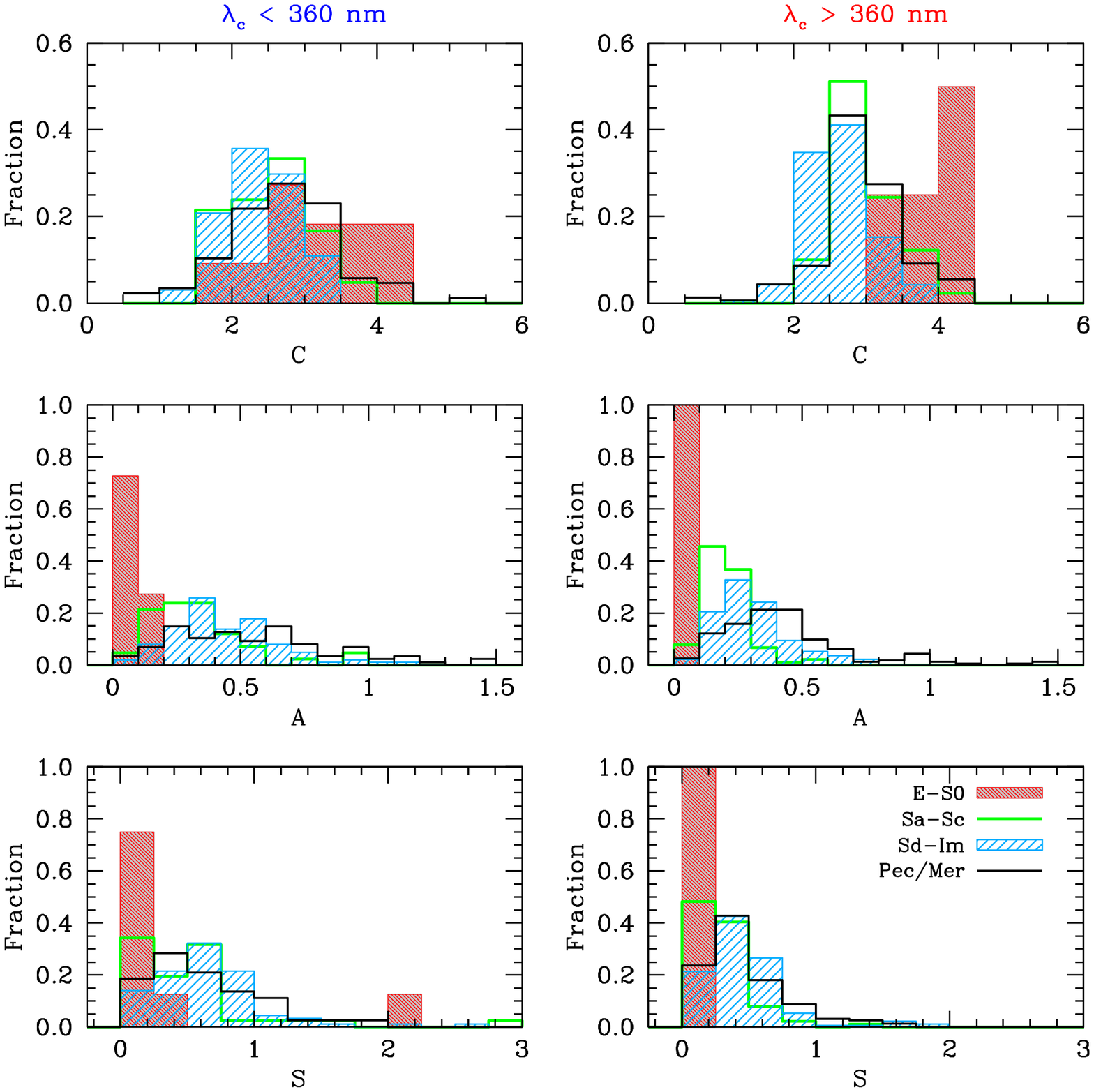}
\figcaption[f10.eps]
{{The distribution of C (upper panels), A (middle panels),
and S (lower panels) for each galaxy type-bin as denoted in the legend
in the lower right panel. The left panels include all UV data from
all filters shortward of the Balmer break ($\lambda_c \lesssim 360$~nm),
and the right panels include all data from all filters longward
of the Balmer break ($\lambda_c \gtrsim 360$~nm). Data that did not meet
our reliability criteria were not included in this plot. Early-type
galaxies (E--S0) have the least amount of overlap with other galaxy
types, particularly in longer wavelengths, and particularly in A and S.
Galaxies of type Sa and later, however, have considerable overlap in
the CAS parameters, although their distributions are offset from
each other. Asymmetry and clumpiness can therefore be used independently
to classify individual early-type galaxies (E--S0) with some limited
degree of certainty, but not individual later-type galaxies. The
overall distribution of the CAS parameter in a large sample, however,
can be used to describe the population distribution as a whole.
}}
\end{figure*}

\subsection{The Distribution of Galaxies in CAS Parameter Space}

The CAS parameters have been shown to correlate with each other to 
form parameter spaces that can be used to classify galaxies, 
distinguish between interacting and merging galaxies, and determine
the extent of recent star formation (Conselice 2003). We examine
the distribution of the galaxies within our sample in the CAS parameter
space by plotting the three parameters against each other in
Figures 11--15.

Figure 11 shows the concentration index for each galaxy as a function of 
asymmetry index and filter. Plot symbols are coded according to galaxy type.
In this figure we do not separate the peculiar and merging galaxies into 
their sub-types. Smaller symbols are used for galaxies that did not
meet our reliability criteria (S/N $< 75$, R $> 1.25$ kpc).
In order to examine the effects of high physical resolution and dust lanes,
we also indicate whether galaxies are particularly nearby (R $< 20$ pc), or
appear to be edge-on spirals. In general, galaxies become less 
concentrated toward 
higher asymmetry, which agrees with the results of other studies
(e.g., Conselice 2003). Galaxy types overlap within this plot, but early-type
galaxies (E--S0) are in general the most concentrated and the most 
symmetric, while later galaxy types become in general less concentrated and
more asymmetric. The merging/peculiar galaxy trend is offset from the normal
galaxy trend, with merging/peculiar galaxies (black symbols) tending to be 
more asymmetric and more concentrated than other galaxy types. The extreme 
outliers from the general normal galaxy trend tend to be fainter than our 
S/N limit of 75, and have large associated measurement uncertainties.
An edge-on orientation did not have a significant effect on the location 
of a galaxy within this parameter space. On average, the more nearby 
galaxies tend to be slightly more asymmetric and less concentrated than 
their more distant counterparts, which results from their partial resolution
into individual stars, as well as a possible selection bias favoring
low luminosity dwarf systems at that distance. No systematic trend was 
noticed in the CAS parameters for galaxies with a physical resolution of 
R $< 20$ pc. 

\begin{figure*}
\plotone{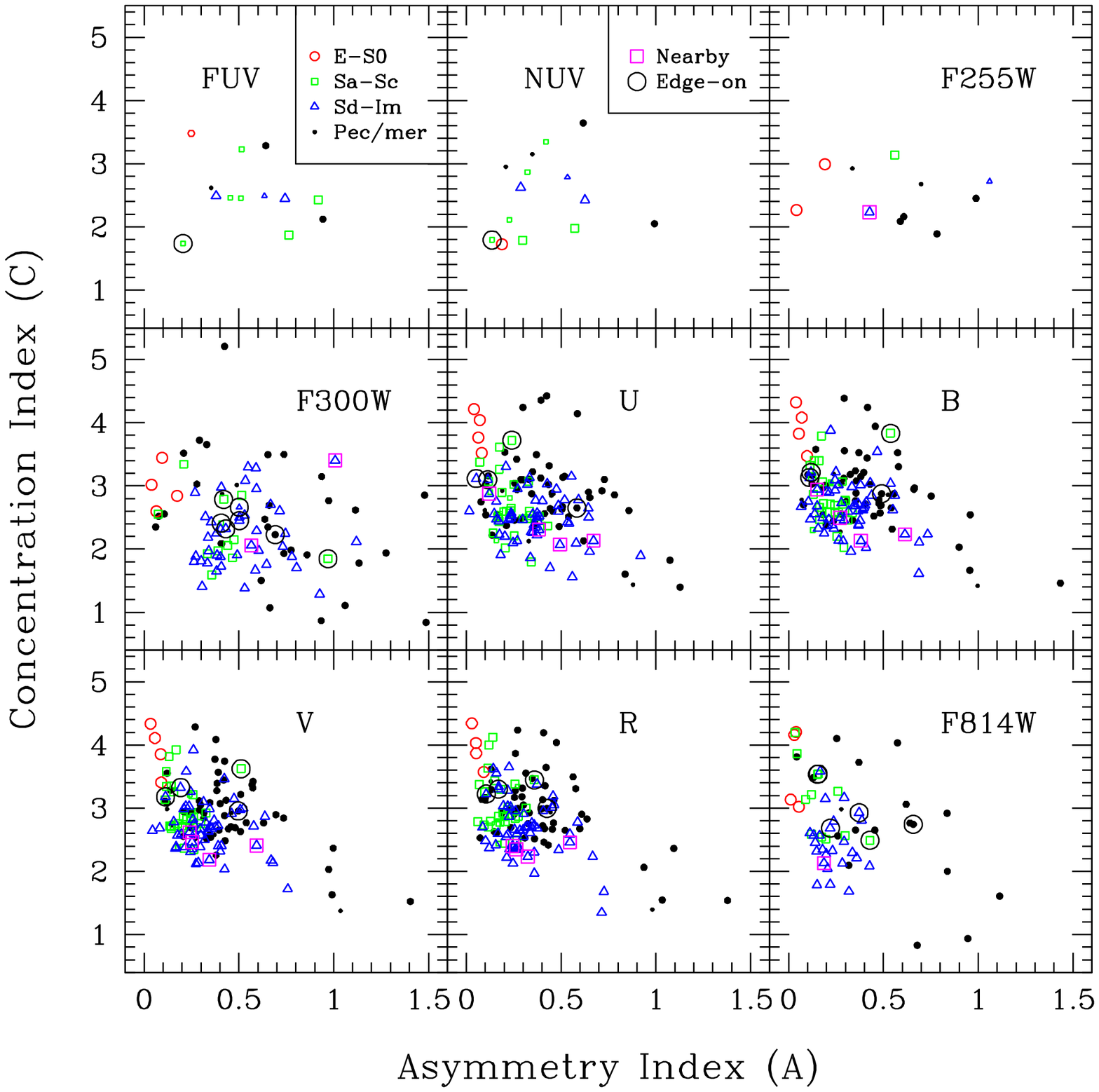}
\figcaption[f11.eps]{{Concentration index as a function of asymmetry
index and filter. Small points represent galaxies that do not meet our
adopted reliability criteria (see text).
Plot symbols are coded according to galaxy type as indicated in the
legend in the upper left (FUV) panel. Here, we do not separate
the peculiar and merging galaxies into their sub-types. We furthermore
highlight nearby galaxies in images with particularly high resolutions
(R $< 20$ pc),
as well as highly inclined/edge-on galaxies, by over-plotting larger symbols
coded according to the legend in the
upper-middle (NUV) panel. In general, galaxies are less concentrated
when they are less symmetric. Increasingly later galaxy types
become increasingly less concentrated and more asymmetric. The locus
of merging/peculiar galaxies is offset from that of the normal
galaxies toward higher asymmetries and concentrations. Extreme outliers
from the general trend are usually fainter than our S/N limit of 75, and have
large associated measurement uncertainties. Particularly nearby galaxies,
which tend to be partially resolved into their individual stars,
are on average slightly less symmetric and less concentrated than
more distant galaxies, due in part to spatial resolution effects.
}}
\end{figure*}

Figure 12 also shows C vs. A for each galaxy, as in Figure 11, but here the
separate sub-classes of peculiar/merging galaxies are represented by
different symbols instead of the normal galaxy type-bins. Although
there is some overlap, peculiar galaxies tend to be more asymmetric and
more concentrated than normal galaxies. For some of these peculiar
galaxies, their high concentration in the UV could be due to the
presence of an AGN. The pre-mergers tend to be 
less asymmetric than the other types of mergers, and follow the general 
trend of normal 
galaxies except that they are on average slightly more asymmetric. This
is to be expected, as the pre-mergers are normal galaxies that are
only beginning to be tidally affected by their neighbors, and may
show some enhanced star formation. There are very few minor mergers
in our sample, with a large scatter in distribution on this plot, so
any conclusions about that sub-class should be regarded with caution. We
only have four major mergers observed in $UBVR$ in our sample, which
is a result of the rarity of these types of galaxies in the nearby
Universe. Although further observations of other mergers are necessary
to improve number statistics, most of the major mergers in our
sample are clearly separated from the main trend, and are much more 
asymmetric and less concentrated than any other galaxy type. 
Merger
remnants, on the other hand, lie closer to the trend line for
normal galaxies, although
they are more asymmetric and more concentrated than both normal galaxies
and pre-mergers. 

\begin{figure*}
\plotone{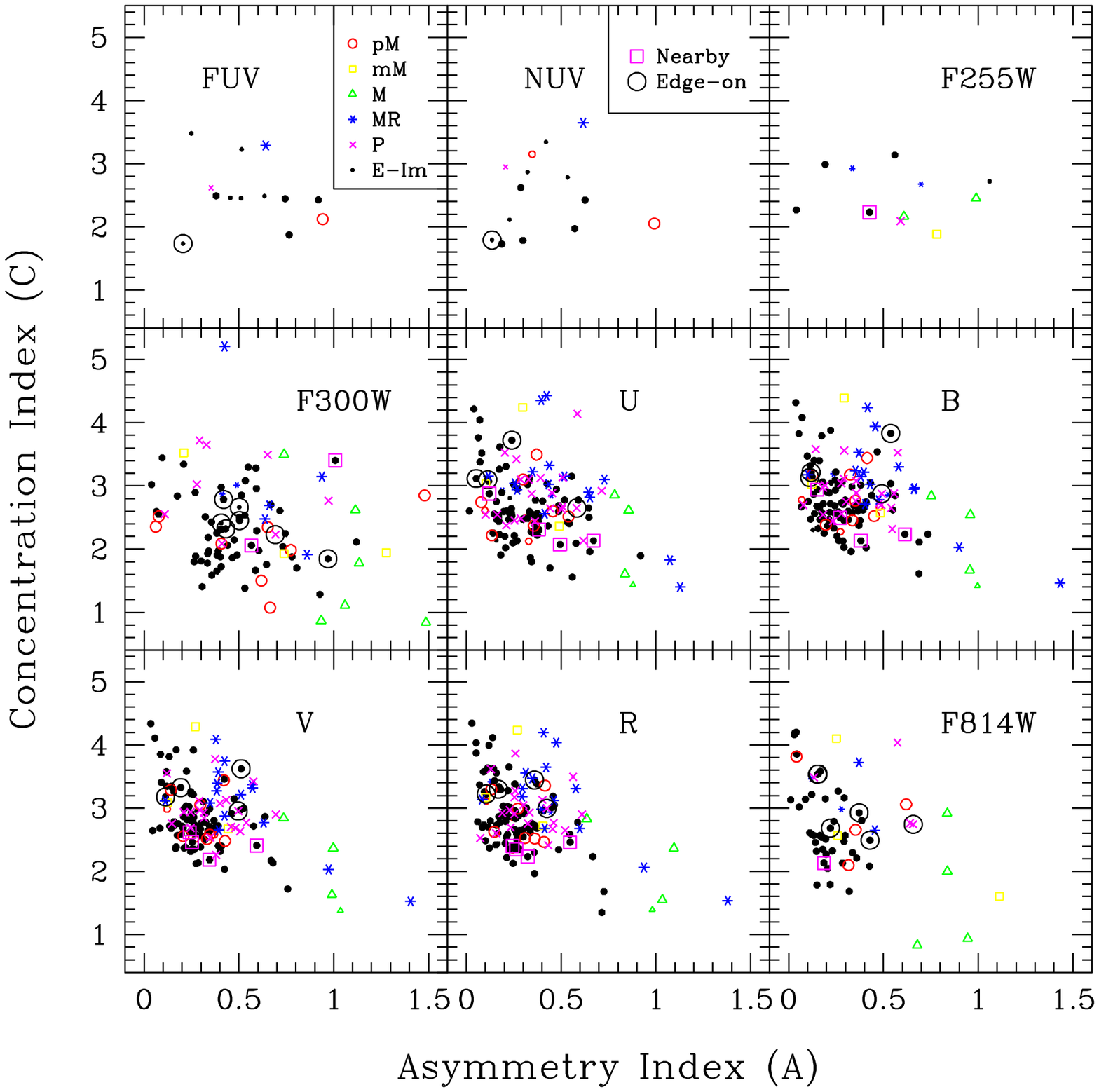}
\figcaption[f12.eps]{{Concentration index vs. asymmetry
index, as in Fig.~11, but with plot symbols highlighting the individual
sub-classes of merging and peculiar galaxies rather than normal
morphological types. Peculiar galaxies (P) are on average less symmetric and
more concentrated than normal galaxies. For some of these peculiar
galaxies, their high concentration in the UV could be due to the
presence of an AGN. Pre-mergers (pM) tend to be slightly
less symmetric than normal galaxies. Major mergers (M) are much less
symmetric
and less concentrated than any other galaxy type. Merger remnants (MR)
are less symmetric and more concentrated than both normal galaxies and
pre-mergers.
}}
\end{figure*}

Figure 13 shows the clumpiness index versus the asymmetry index for 
each galaxy, using the same normal galaxy type symbols
as in Figure 11. Measurements of S $< 0$ can occur for particularly
smooth or faint galaxies where the sky was measured to be more clumpy 
than the galaxy itself. This is more likely to occur with HST/WFPC2, due
to large noise spikes caused by local CTE effects. The degree to which
these S measurements lie below zero is within the uncertainties on S,
however. Galaxies tend on average to be more clumpy with higher
asymmetry, which agrees with the results of other studies (e.g., 
Conselice 2003). This relation is fairly tight, especially at longer 
wavelengths,
as S and A are not entirely independent from each other. Early-type
galaxies (E--S0) are the least asymmetric and clumpy, with later galaxy
types becoming progressively more asymmetric and more clumpy. 
Outliers from the general trend in the shorter wavelengths have particularly
low S/N in the images, while outliers in the longer wavelengths are 
either edge-on or are very nearby (with resolution, R $< 20$ pc).
Due to the strong dust lanes visible in edge-on galaxies, 
they appear more clumpy than galaxies with lower inclinations, and 
lie well above the general S vs. A relation. On the whole, 
nearby galaxies tend to appear to be only slightly more 
asymmetric and clumpy than their more distant counterparts.

\begin{figure*}
\plotone{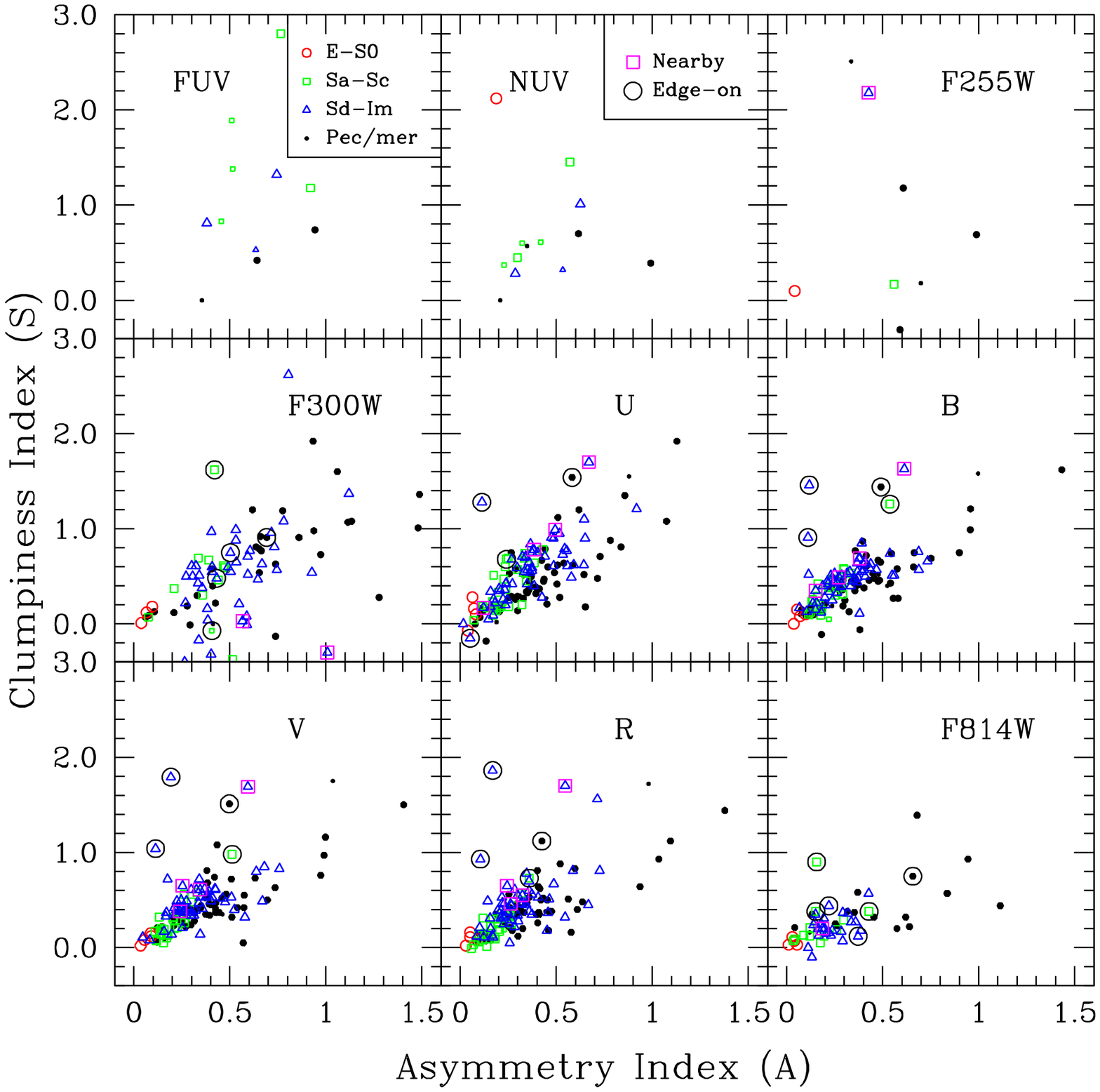}
\figcaption[f13.eps]{{Clumpiness index vs. asymmetry index.
Symbols are coded as in Fig.~11. Galaxies are on average more clumpy
when they are less symmetric. Increasingly later galaxy types become
progressively less symmetric and more clumpy. Outliers in the shorter
wavelengths tend to have S/N $< 75$, and large
associated uncertainties.
Edge-on spirals are well separated from the general trend, appearing
more clumpy than galaxies with lower inclinations. This is likely due to
more visible dust-lanes that increase the clumpiness index.
Particularly nearby galaxies
tend to be slightly less symmetric and more clumpy than more distant galaxies,
due to resolution effects.
}}
\end{figure*}

Figure 14 also shows S vs. A for each galaxy, but with peculiar/merging 
sub-classes highlighted with different symbols, as in Figure 12. The
pre-mergers are not clearly distinguished from normal galaxy types
within the measurement uncertainties, but
merging galaxies are much more asymmetric and show much more small
scale structure (clumpiness) than any other
galaxy type. Merger remnants tend to appear slightly smoother, but 
more asymmetric on average than the pre-merger and normal galaxies.

\begin{figure*}
\plotone{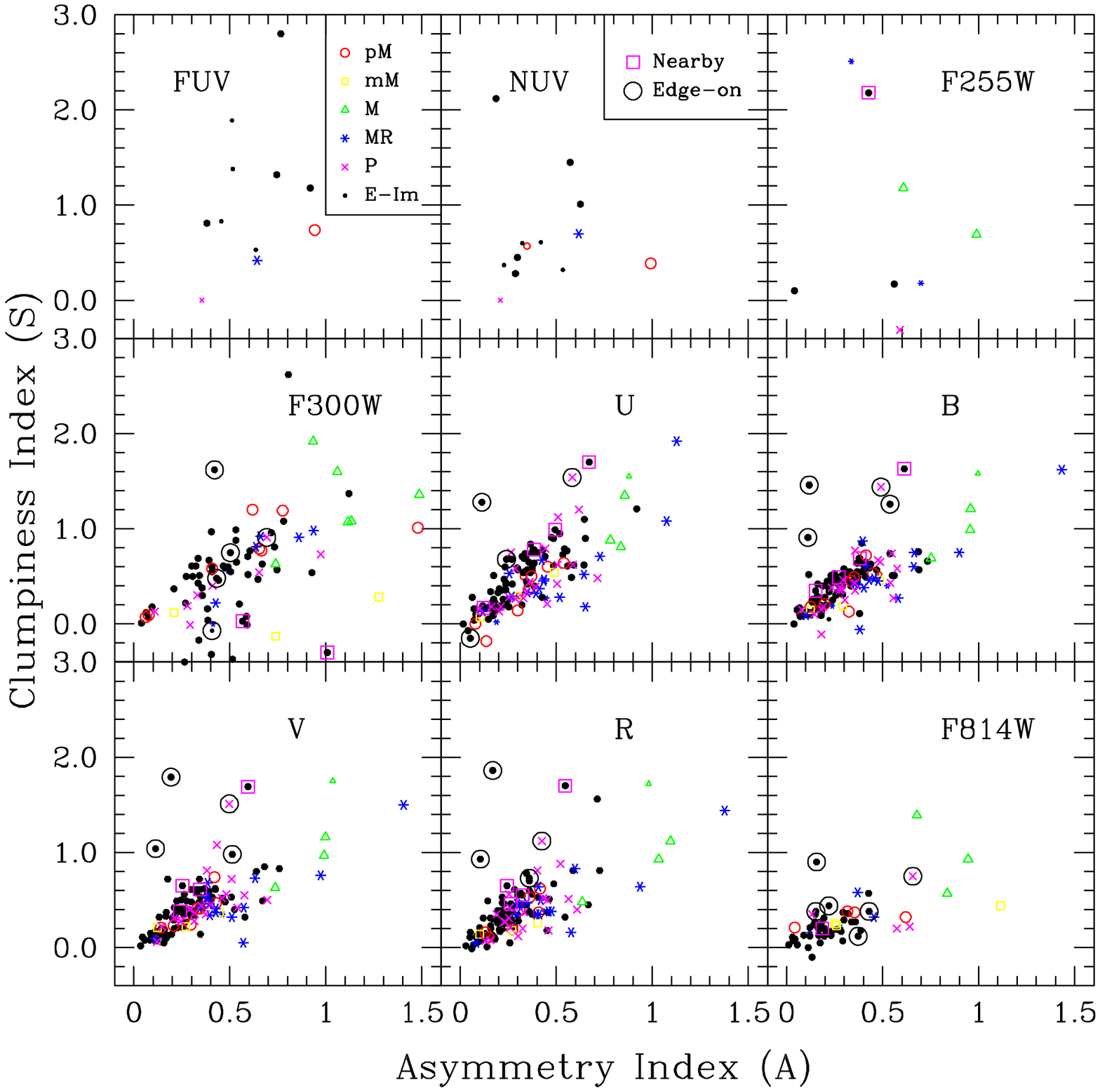}
\figcaption[f14.eps]{{Clumpiness index vs. asymmetry index, with
merging and peculiar galaxies separated into their sub-classes. Symbols are
coded as in Fig.~12. Merging galaxies are much less symmetric and more
clumpy than any other galaxy type. Merger remnants tend to be slightly
less clumpy and less symmetric than both pre-mergers and normal galaxies.
}}
\end{figure*}

In Figure 15, we examine the average trends in this CAS parameter space
with a plot of the average of the median values in each type-bin and 
filter of C and S as a function of A. The left panels contain the average
of the median values in each filter shortward of the Balmer break 
($\lambda_c \lesssim 360$ nm). The right panels contain the average of the 
median values in each filter longward of the
Balmer break ($\lambda_c \gtrsim 360$ nm). Galaxies that did not meet our
reliability criteria were not included in the medians. Type-bins in
each filter with fewer than 2 galaxies were not used in the average.
The error bars are not the errors on the averages, but rather the
average value of the 25--75\% quartile ranges for each filter used
to calculate the average. Therefore, the error bars give an indication
of how much an individual galaxy may vary in this parameter space from
the average value. 
There is a much larger scatter in the CAS parameters at shorter
wavelengths, which may partially be due to larger measurement 
uncertainties for these
filters and smaller number statistics. More, and higher quality
data in the mid to far-UV is needed to further constrain these values. 
There is a clear trend among normal galaxies to show a progression 
from early to late-type galaxies (E to Im) of increasing asymmetry and 
clumpiness and decreasing concentration, particularly in the redder
filters.
The different peculiar/merging galaxy types are each located at different
positions within this parameter space. There appears to be a general
evolutionary trend from pre-mergers to mergers, and then to post-mergers, 
as shown by the dotted lines with arrows connecting the average of the 
median values 
for each type-bin. Mergers become significantly less concentrated,
more asymmetric, and more clumpy than pre-mergers, then progress
back toward the normal galaxy parameter space as they turn into merger
remnants, which end up being more concentrated than
the pre-mergers, and slightly more asymmetric and clumpy. From there,
galaxies may take different paths on the CAS parameter space depending
on whether they turn into elliptical or spiral galaxies, which
depends on the details of the merger and of the individual galaxies
taking part in that merger. 

\begin{figure*}
\plotone{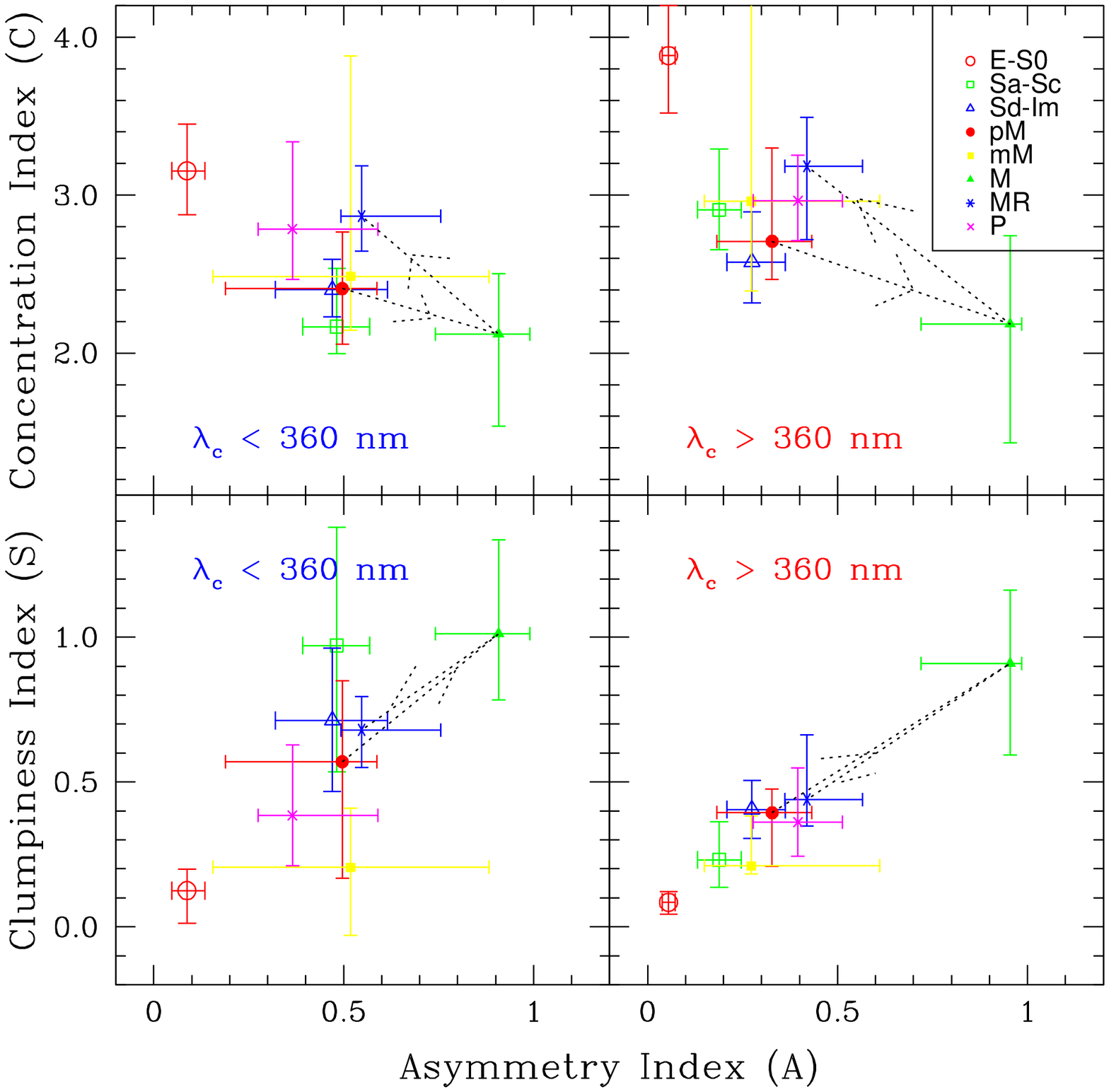}
\figcaption[f15.eps]{{The median values of the
concentration and clumpiness indices for each type
bin vs. the median values of the asymmetry index in that
type-bin. Symbols are coded by type as described in the upper right
panel. {\it Left panels:} The average of the median values in each
filter short-ward of 
the Balmer break ($\lambda_c \lesssim 360$ nm). {\it Right panels:} The
average of the median values in each filter long-ward of the
Balmer break ($\lambda_c \gtrsim 360$ nm).
Galaxies that did not meet our reliability criteria were not included
in the medians. Type-bins in each filter with fewer than 2 galaxies
were not used in the average. The error bars show the 25--75\% quartile
ranges of the
CAS parameters, averaged over each filter used in the average.
There is a larger scatter in the CAS parameters at shorter
wavelengths, which may partially be due to lower galaxy number statistics
and lower S/N ratio within the GALEX and HST UV images.
Normal galaxies become more asymmetric and clumpy and less
concentrated toward later Hubble type (E to Im).
Dotted lines connect the average of the medians for pre-mergers to mergers,
and then to post-mergers. This shows a progression through the CAS
parameter space as the merging process progresses,
with mergers becoming significantly less concentrated, more
asymmetric, and more clumpy than pre-mergers, before returning as
post-mergers near
the locus of normal galaxies again. On average the merger remnants are
more concentrated than the pre-mergers, and slightly
more asymmetric and clumpy. This may define a duty cycle for the
merging process.
}}
\end{figure*}

\subsection{Quantitative Classification Using CAS in the Optical and 
Ultraviolet}

One of the major goals of quantitative morphological studies is to 
automatically distinguish galaxy types from each
other. This has proven to be a difficult
problem, and various pieces of information, such as colors, magnitude
and sizes, are needed in addition to the CAS parameters to fully classify 
galaxies into various Hubble types automatically. No
perfect method has yet been developed. We can however determine
how well basic CAS classification boundaries do in terms of
automatically finding galaxies of various types, without any other
information involved.  

A revised CAS classification space can be defined as follows for
finding galaxies with early, mid, late and merger/peculiar morphologies
in rest-frame optical images (see Bershady et al. 2000). Early-types 
are systems with A $< 0.1$ and typically C $>3.5$, mid-types are those
with $0.1 < $ A $< 0.35$ and C $> 2.44\times{\rm log A} + 5.49$, late
types with $0.1 <$ A $< 0.35$ and C $< 2.44\times{\rm log A} + 5.49$, and
finally mergers have A $> 0.35$ and A $> S$ (Conselice et al. 2004).
To carry out the CAS reliability test we examined the position in
CAS space of each morphology type to determine what fraction of eye-ball
estimates of morphology match the CAS classification boundaries, that is
the fraction that are correctly matched (f$_{\rm correct}$). We also measure
the fraction of galaxies within each CAS type volume which do not belong
to the assigned type, which we call the contamination fraction
(f$_{\rm cont}$). We remove the nearby galaxies from this analysis to avoid
resolution issues.

It turns out that the best results are within the optical region, where
the CAS system was originally designed to work. We find that
for early-types, f$_{\rm correct} \sim$ 1 in the $UBVR$ and F814W bands.
The F225W, NUV and FUV
bands only have a few early-types which are too faint for a reliable
CAS measurement, but their asymmetry values tend to be a bit higher.
The contamination is also very low with f$_{\rm cont}$ = 0 for early-types
in the UBVR bands.  Most of the contamination in the optical within the
A $< 0.1$ CAS space is largely due to mid-types.
However, we only have a few early-types and
mid-types in our sample, and we are mainly
interest in distinguishing late-types from mergers/peculiars,
as these galaxies dominate the high redshift population.

The separation of mergers from late-type galaxies can be successfully done
within CAS space, although there is some overlap. The particular problem, 
first stated in Conselice (2003), is that not all peculiar/mergers follow 
the CAS space definition of high asymmetry. Conselice (2003) made this
conclusion based on the location of 66 Ultra-Luminous Infrared Galaxies
(ULIRGs) in CAS space. About half met the criteria A $> 0.35$, while the
remainder were at lower asymmetry, with the systems spread out amongst
the concentration index. Using the standard criteria listed above,
we find that for late-types, f$_{\rm correct} \sim 0.5-0.9$ in the $UBVR$ and
F814W bands, with the lowest fraction f$_{\rm correct} = 0.5$ for the U
band, and the highest f$_{\rm correct} \sim 0.9$ in F814W. Using the same CAS
criteria for finding late-types, we find that the contamination ranges
from 50\% in the U-band, to 20\% in the F814W band, and about 40\% in the
BVR bands. A large part of this contamination arises from peculiar
and merging galaxies, which have similar asymmetries. However 70\% of
galaxies in the late-type regime are classified as late-type by eye, which
is a reasonable success rate.

The fraction of mergers found with the basic CAS criteria is not as high,
with  f$_{\rm correct} \sim 0.3-0.4$ in the $UBVR$ bands, and 0.6 in
the F814W band. The contamination is also fairly low, with
f$_{\rm cont} \sim$ 0.2. This result shows that while most galaxies
with a high asymmetry in the UBRI bands will be mergers/peculiars, not
all mergers/peculiars have large asymmetries (see Conselice 2003).
Likewise, some high-A galaxies, particularly in the bluer bands are not
classified as a merger or peculiar.  It is worth noting that the
best results are obtained using the reddest band, F814W, which suggests
that measurements of galaxy structure are best done in the reddest
wave-band possible.

A major goal of this paper is to test whether we can tell the difference
between late-type star forming galaxies and mergers through the CAS
parameters in the ultraviolet. It turns out that the ability to distinguish
late-types from mergers/peculiars is slightly worse in the UV than in the 
optical, but not by much.
Using the  standard CAS method, 50\% of eye-ball classified late-types
fall in their correct CAS region in the NUV.
The late-types in the UV which do not fall in the correct UV-CAS region
typically have high asymmetries, A $> 0.35$ (Figure 11). Generally, these
galaxies have high clumpiness values as well. We can use this fact to
better automatically separate late-type galaxies from mergers/peculiars
in the ultraviolet. If we take the following criteria:

\begin{equation}
A > 0.25 {\rm ~and~} C < 2.44\times{\rm log}A + 5.49 {\rm ~and~} S > A,
\end{equation}

\noindent we acquire 60\% of the late-types, with a 40\% contamination rate
from mostly peculiars/mergers in the F300W and U-bands. This contamination
is again largely the result of peculiar/mergers.

Using the standard revised optical CAS method for finding mergers (A $> 0.35$
and A $> S$) we find that all mergers/peculiars are identified in the FUV,
50\% in the NUV, 75\% in the F225W band and 40\% in the F300W band.
The contamination rates within the area of CAS for mergers in the
FUV, NUV, F225W and F300W are 0\%, 0\%, 25\%, and 33\% respectively.
It is clear from Figures 11--14 that the asymmetry limit of
A $> 0.35$ has a contamination from late-type galaxies. This is
due to the increased prominence of star formation in the UV, creating
high asymmetries which are sensitive to star formation and merging
activity (Conselice et al. 2000).
However, if we increase the limit to A $= 0.6$ in the UV, we are
able to obtain a nearly clean sample of mergers. This new UV-CAS criteria
for finding mergers is:

\begin{equation}
A > 0.5 {\rm ~and~ } A > S.
\end{equation}

\noindent Using equation (5) to find mergers in the F300W band we get
f$_{\rm correct} \sim$ 0.4, while  f$_{\rm con} = 0.25$.
In the F255W band the contamination is also low, f$_{\rm con} = 0.25$, and
the f$_{\rm correct} \sim$ 0.75, although there are too few merging galaxies
(3) in this pass-band to draw definite conclusions. The NUV and FUV
bands have too few galaxies to reliably compute these statistics.
We suggest that future studies that use rest-frame
UV morphologies of high redshift galaxies use the UV-CAS equations (4--5) to
separate late-types from mergers, although larger number statistics in the
UV are needed to verify these results.

\subsection{The Rest-frame Wavelength Dependence of the CAS Parameters.}

Some of the galaxies in our sample appear different in the UV than they
do in the optical, while some galaxies appear very similar. This
was discussed qualitatively in detail in Windhorst et al. (2002).
Figure 16 shows
examples of each of these cases. The upper image for each galaxy was
obtained with either the F814W near-IR filter, or the VATT R-band filter.
The lower image for each galaxy was obtained with either the F300W
UV filter, or the VATT U-band filter. UGC05189 is a merger remnant that
has a large difference in asymmetries and clumpiness indices between the
UV and the IR (A$_{F300W}$ -- A$_{F814W}$ = 0.182, and
S$_{F300W}$ -- S$_{F814W}$ = 0.49), mostly due to bright star-forming knots
that are more apparent in the UV image. UGC06816
is a late-type barred spiral galaxy that has a large difference
in A and S between the near-UV and the red (A$_{U}$ -- A$_{R}$ = 0.323, and
S$_{U}$ -- S$_{R}$ = 0.64), which is
also due to its UV-bright star-formating knots, particularly at
the ends of its bar and at two sites of vigorous star formation
along its poorly organized spiral pattern. Both of these galaxies
have small differences in concentration index between these filters
(--0.178 for UGC05189, and --0.207 for UGC06816). UGC12808
is an earlier-type spiral galaxy that has about the same asymmetry and
clumpiness index in both pass-bands (A$_{F300W}$ -- A$_{F814W}$ = --0.045,
and S$_{F300W}$ -- S$_{F814W}$ = --0.05). UGC06697 is a peculiar, highly 
inclined galaxy that also has about the same asymmetry and clumpiness
index in both pass-bands (A$_{F300W}$ -- A$_{F814W}$ = 0.035, and
S$_{F300W}$ -- S$_{F814W}$ = 0.16). Both of these galaxies, however,
appear different in the UV than in the IR due to a large difference
in concentration index between the filters (--0.664 for UGC12808 and
--0.515 for UGC06697), which is largely due to the diminished appearance of
their red bulges at shorter wavelengths. A quantitative
discussion of the dependence of the CAS parameters on rest-frame
wavelength and galaxy type could serve as a zero-redshift benchmark for
higher redshift galaxy classifications. As such, we present the results for
our dataset below.

\begin{figure*}
\plotone{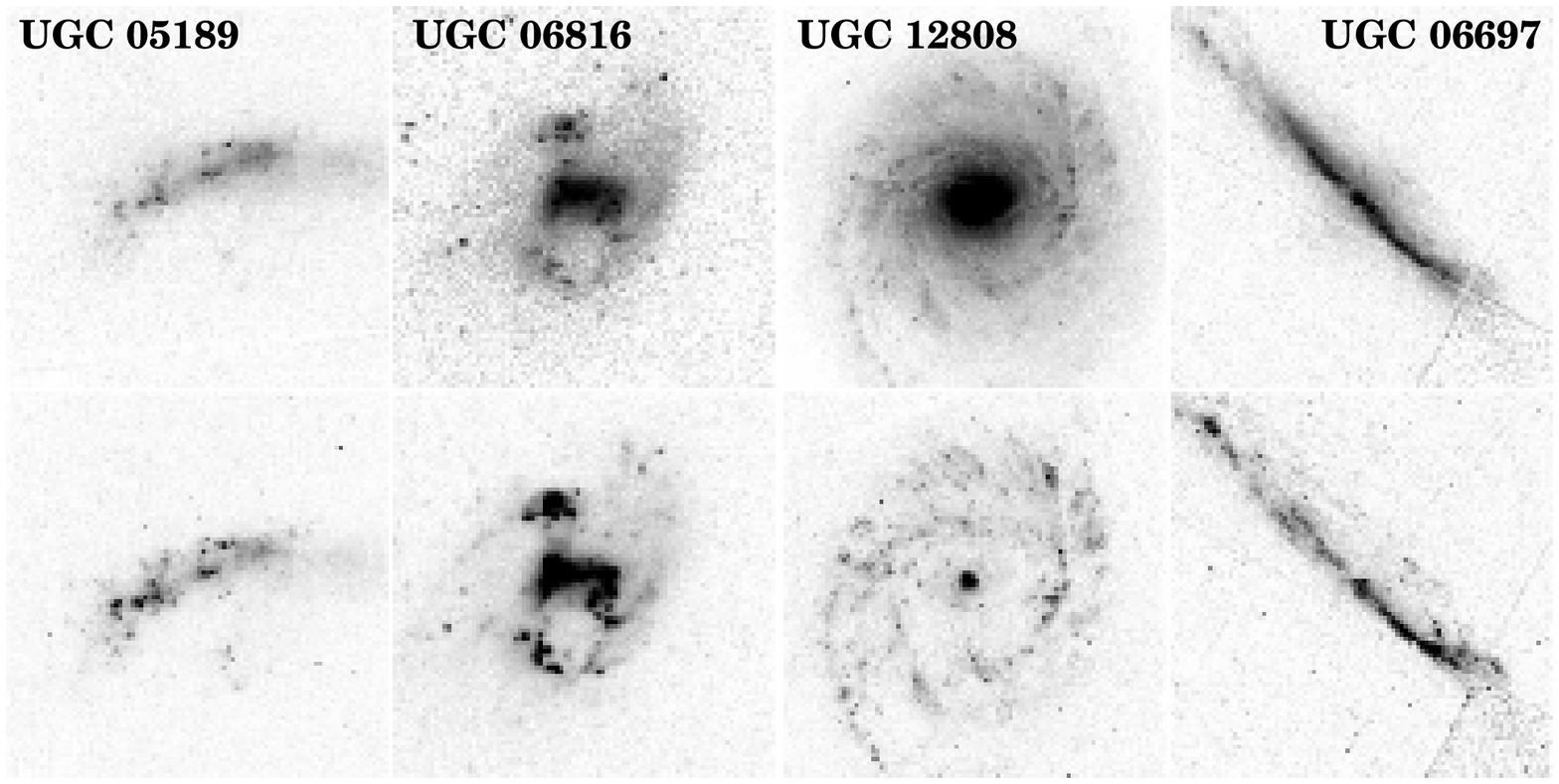}
\figcaption[f16.ps]
{{Examples of galaxies that have various differences
in the CAS parameters as measured in the UV vs. the optical/near-IR. The
upper panels are the red (R-band) or near-IR (F814W) images for each
galaxy. The lower panels are the UV (U-band or F300W) images for these
galaxies. From left to right: UGC05189 is a merger remnant with a large
difference in A and S between the UV and the near-IR, but little difference
in C. UGC06816 is a late-type spiral galaxy that also has a large
difference in A and S between the near-UV and the red, but little difference
in C. UGC12808 (NGC7769) is an early-type spiral galaxy that has little
difference in A and S between the UV and the near-IR, but a large difference
in C. UGC06697 is a peculiar galaxy that also has little difference in
A and S between the UV and the near-IR, but a large difference in C. See
the text for the differences in CAS parameters for each of
these galaxies.
}}
\end{figure*}

\begin{figure*}
\plotone{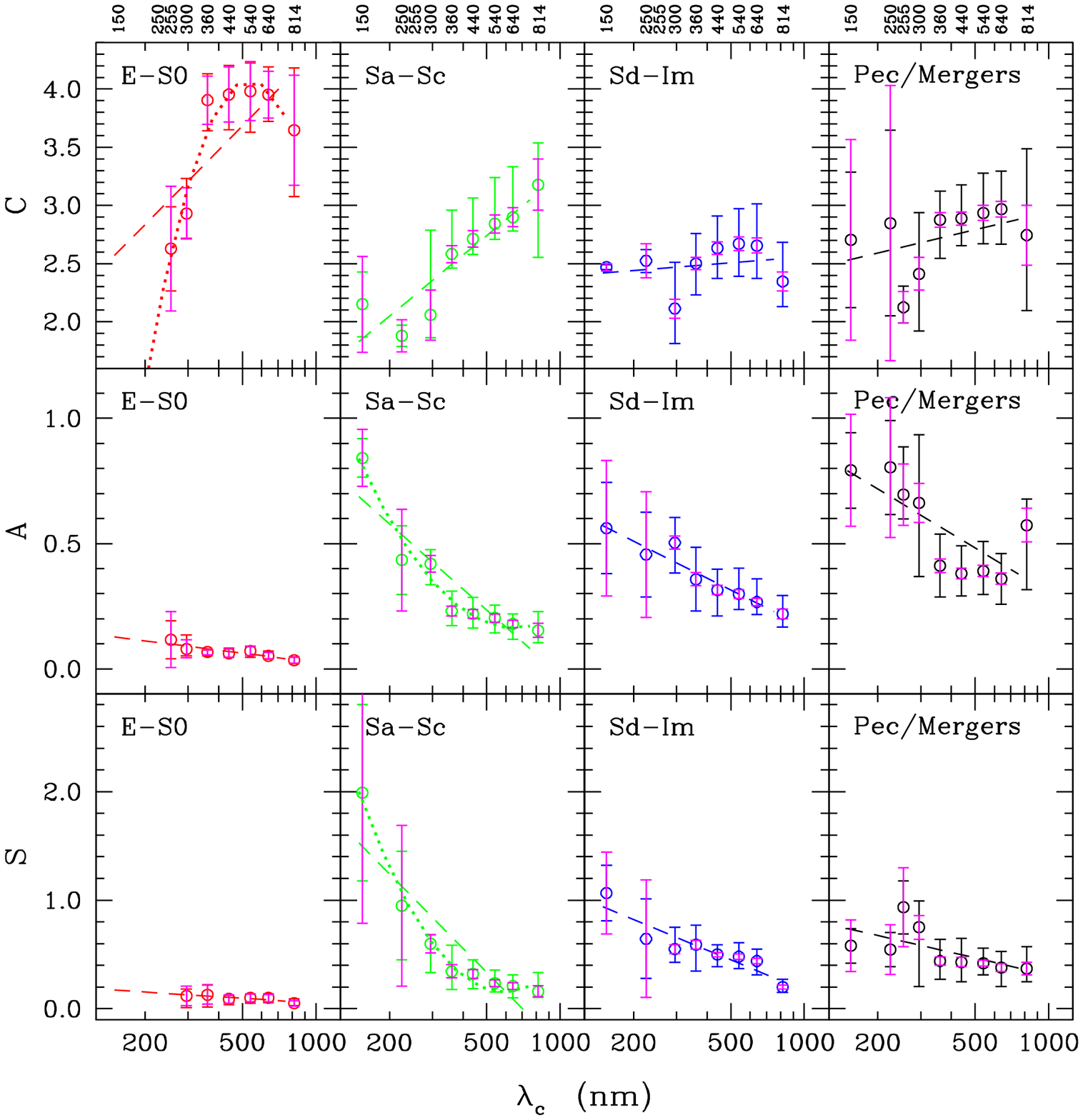}
\figcaption[f17.eps]
{{The median CAS parameters as a function
of the central wavelength in nm ($\lambda_c$) of each filter.
The values for different type-bins are separated into different panels,
from left to right. Error bars with the same color as the data points
show the 25--75\% quartile ranges. Magenta error bars show the error
on the median.
Data that did not meet our reliability criteria, and CAS values $<-0.2$
are not included in the medians. Parameters for type-bins and filters
with fewer than 2 galaxies are not plotted. Dashed lines are the
linear-least-squares fit to the median CAS parameters. Dotted
lines are the 2nd order polynomial fits to the data, shown only in
the C panel for E--S0 galaxies, and the A and S panel for Sa--Sc galaxies.
These may be the only panels where the trend is of higher order than linear.
These fits show an increase in C, and a decrease in A and S with
increasing rest-frame wavelength, although early-type galaxies (E--S0)
are relatively constant in C, A, and S with increasing rest-frame
wavelength longward of the Balmer break. Shortward of the Balmer break,
the trend for early-type galaxies is more uncertain due to
the low S/N of these intrinsically red galaxies in the UV images.
}}
\end{figure*}

Figure 17 shows the median values of the CAS parameters for each type-bin
as a function of the central wavelength ($\lambda_c$) of each 
filter in nm. Values of the CAS parameters that were $< -0.2$ were not
included in the median, as well as values from galaxy images that did 
not meet our reliability criteria. We do not include medians in this plot 
for filters and type-bins that have fewer than 2 galaxies. Error bars
that are the same color as the data points indicate the 25--75\% 
quartile range. Magenta error bars indicate the error on the median, which
is smaller for bins that contain larger number statistics. We find a 
large scatter among 
individual galaxies, but there is a general trend of increasing C and 
decreasing A and S toward longer wavelengths. As in previous 
plots of the CAS parameters, on average, the early-type, late-type, and
merging/peculiar galaxy classes are clearly separated in C, A, and S, 
although there is considerable overlap among individual galaxies
within each type class. The CAS values of early-type galaxies (E--S0) 
are relatively constant with increasing rest-frame wavelength longward of the
Balmer break. Shortward of the Balmer break, the S/N of
these red objects is very low, and therefore the CAS parameter measurement
uncertainties are high. Due to this effect and due to
the lower number of early-type galaxies in our sample, we cannot 
draw definite conclusions about whether the CAS parameters of
early-type galaxies would differ significantly at UV rest-frame
wavelengths. Spiral, irregular,
and peculiar/merging galaxies, however, all show a general trend of 
increasing C and decreasing S and A with increasing rest-frame
wavelength. The concentrations measured in the HST/WFPC2 F814W images
appear somewhat lower with respect to the general trend,  
which may be due to the fact that our WFPC2 images are more
sensitive to point-sources than extended sources, so that outlying low
surface-brightness material may not have been detected in that particular 
filter.

Also, in Figure 17, linear-least-squares fits to the median data points 
are shown as dashed lines. Linear fits appear suitable for all but the
wavelength dependence of the concentration index of early-type (E-S0)
galaxies (which shows more sharply decreasing C measurements 
at wavelengths shortward of the Balmer break), and the asymmetry and
clumpiness indices of mid-type (Sa-Sc) galaxies (which show more
sharply increasing A and S measurements shortward of the Balmer break). 
Therefore, we also include a 2nd order
polynomial fit to the data in these panels, shown as dotted lines. This
strong dependence of C shortward of the Balmer break for the E--S0 galaxies
may be mostly due to the very low S/N of these red galaxies in the UV 
images (see Windhorst et al. 2002 for the images).
This 2nd order polynomial is given by:
\begin{equation}
C(E-S0) = -14.33~log(\lambda_c)^2 + 78.21~log(\lambda_c) - 102.6
\end{equation}
\noindent This is a relatively good fit to the data, with
~13\% uncertainty on each of the coefficients (compared to a 39\% error on
the slope of the linear fit, and an error larger than the value of the
y-intercept). The strong dependence of A and S shortward of the Balmer
break for the Sa--Sc galaxies may be due to UV-bright star-forming regions,
although the low number statistics in the bluest two filters (2 galaxies)
and the large uncertainties suggest that this may be insignificant.
However, the general trend of larger A and S values at shorter wavelengths
is likely real, as the F300W filter contains 11 Sa--Sc galaxies, and also
shows this trend. More far-UV data is needed to more accurately determine 
the trend in this wavelength range.
The 2nd order polynomials for the dependence of A and S
on wavelength for Sa--Sc galaxies are: 
\begin{equation}
A(Sa-Sc) = 1.69~log(\lambda_c)^2 - 9.48~log(\lambda_c) + 13.47
\end{equation}
\begin{equation}
S(Sa-Sc) = 5.20~log(\lambda_c)^2 - 28.82~log(\lambda_c) + 40.11
\end{equation}
\noindent The uncertainties on these coefficients are $\lesssim 18\%$, and 
$\lesssim 10\%$, respectively, compared to an error on the slope of the 
linear fits of 17\% and 18\%, respectively.

The coefficients for the linear-least-squares fit are given in
Table 9, where X = X$_1 \times$ log($\lambda_c$) + X$_2$, and where 
$\lambda_c$ is in units of nm. The variable X is the concentration index 
(C for columns 2--3), the asymmetry index
(A for columns 4--5), or the clumpiness index (S for columns 6--7).
As determined from these fits, the early to mid-type spiral galaxies
(Sa--Sc) show a significant increase in concentration index at longer 
rest-frame wavelengths. The late-type (Sd--Im) galaxies, however, have 
no trend with wavelength within the uncertainties, 
and there is only a slight trend with wavelength for 
peculiar/merging galaxies. 
There is, however, a significant decrease in asymmetry toward longer 
wavelengths for all galaxy types. 
The clumpiness index also decreases significantly with wavelength for
all galaxy types. The reader should
be advised that the linear fits of the wavelength dependence of A and S for
Sa--Sc galaxies do not fit the data as well as the 2nd order polynomial 
fits presented in Eq's 7--8. 

%
\begin{deluxetable}{lrrrrrr}
\tabletypesize{\scriptsize}
\tablecolumns{7}
\tablewidth{0.995\columnwidth}
\tablecaption{CAS Wavelength Dependence}
\tablehead{
\colhead{Type-bin} & \colhead{C$_1$} & \colhead{C$_2$} & \colhead{A$_1$} &
\colhead{A$_2$} & \colhead{S$_1$} & \colhead{S$_2$}  \\
\colhead{(1)} & \colhead{(2)} & \colhead{(3)} & \colhead{(4)} &
\colhead{(5)} & \colhead{(6)} & \colhead{(7)}
}
\startdata
 E--S0 &  \nodata &  \nodata & -0.12 & 0.40 & -0.14 & 0.48 \\
  \nodata &  \nodata & \nodata & 0.02 & 0.07 & 0.04 & 0.10 \\
 Sa--Sc &  1.74 &  -1.96 & -0.87 & 2.57 & -2.28 & 6.50 \\
  \nodata &  0.28 & 0.73 & 0.15 & 0.38 & 0.42 & 1.09 \\
 Sd--Im &  0.17 &  2.05 & -0.49 & 1.63 & -0.94 & 2.97 \\
  \nodata &  0.27 & 0.70 & 0.05 & 0.13 & 0.14 & 0.36 \\
 Pec/merger &  0.51 &  1.43 & -0.59 & 2.08 & -0.52 & 1.87 \\
  \nodata &  0.36 & 0.94 & 0.17 & 0.44 & 0.21 & 0.55 \\
\enddata
\tablecomments{Coefficients for the linear dependence of the CAS
parameters on the logarithm of the wavelength in nm. The second line
for each type-bin lists the uncertainties on these coefficients.
The coefficients
are given for the linear-least-squares fit to the data within each
type bin, resulting in the equation: X = X$_1 \times$ log($\lambda_c$) + X$_2$,
where X is the concentration index (C) for columns 2--3, the asymmetry
index (A), for columns 4--5, and the clumpiness index (S) for columns
6--7. Note that better polynomial fits for C (E--S0), and for A and S (Sa--Sc)
are given by Eq's 6--8 in the text.
}
\end{deluxetable}

Because the same galaxies were not observed at all wavelengths,
it is more meaningful to determine the dependence of the CAS parameters
on wavelength by examining the differences seen within individual galaxies,
rather than fitting trends to the entire data set.
Therefore, we also calculated the difference between the CAS parameters 
in $R$ and $U$ for each individual galaxy that met our reliability criteria, 
and which have CAS values $> -0.2$. We use these filters because we have 
higher number statistics in the ground-based VATT data, and all of the 143 
VATT galaxies were consistently observed in $U$ and $R$, but not necessarily 
in the HST and GALEX filter sets. We find median differences between C in 
the $R$- and $U$-band (C$_R$ -- C$_U$) of 0.081, 0.378, 0.139, and 0.135, 
for E--S0, Sa--Sc, Sd--Im, and peculiar/merging galaxies, respectively. 
A positive number indicates an increase in the index at longer wavelengths. 
This difference in C is significant within the scatter for individual 
galaxies (defined as half of the 25--75\% quartile range of C$_R$ -- C$_U$) 
for E--S0 and Sa--Sc galaxies, which have a scatter of $\pm$
0.048 and 0.121, respectively. This is less significant
for later-type galaxies, however as there is a larger range of possible
C$_R$ -- C$_U$ values for individual galaxies (with a scatter of $\pm$
0.162 for Sd--Im, and 0.152 for peculiar/mergers). The asymmetry index
only varies significantly within the spread for E--S0 galaxies,
with a small decrease in A from $U$ to $R$ (A$_R$ -- A$_U$ = --0.011), 
and a scatter in A$_R$ -- A$_U$ of $\pm 0.008$.
The difference between A in $R$ and $U$ is larger for the later-type
galaxies, but so is the spread in values between individual galaxies.
A$_R$ -- A$_U$ $\pm$ the scatter in A$_R$ -- A$_U$ for Sa--Sc,
Sd--Im, and mergers, respectively is: $-0.057 \pm 0.059$, 
$-0.064 \pm 0.078$, and $-0.044 \pm 0.071$. The clumpiness parameter
also decreases at longer wavelengths, although the scatter is large.
S$_R$ -- S$_U$ $\pm$ the scatter in S$_R$ -- S$_U$ for E--S0,
Sa--Sc, Sd--Im, and mergers, respectively is: $-0.030 \pm 0.032$,
$-0.160 \pm 0.130$, $-0.140 \pm 0.195$, and $-0.065 \pm 0.160$. So, 
even though there is a general increase in C and decrease in A and S
toward longer wavelengths, the degree to which this difference occurs
varies significantly between individual galaxies. The values presented
here may be used as a correction to the general statistics of a large
sample of galaxies, but cannot be applied with a large degree of 
certainty to any one individual galaxy.

\section{Conclusions}

Concentration, asymmetry, and clumpiness (CAS) indices are basic
structural parameters of galaxies.
Because the CAS parameters are based on the light distributions
of galaxies, they describe quantifiable aspects of their morphologies.
A difference in these parameters as a function of  
wavelength therefore contains information about how a galaxy's 
appearance changes with rest-frame wavelength, and
leads to a measure of the morphological k-correction. This is
important for high redshift studies, because at high redshift these
galaxies are observed in their rest-frame UV. To quantify this, we
have presented CAS parameter measurements for our sample of 199 mostly
late-type and merging/peculiar 
galaxies observed in various combinations of filters spanning the far-UV
through the near-IR.

Our results confirm those of previous studies, but over a much wider
wavelength range that includes the UV: galaxies
are more concentrated, less asymmetric, and less clumpy with redder
color and earlier galaxy type. We find that normal and merging/peculiar 
galaxy type classes occupy different locations within the CAS parameter 
space, such that correlations between the CAS parameters
can be used for statistical galaxy classification, but in a limited 
sense due to both intrinsic and observational scatter. 

A general evolutionary trend as a merger progresses through its different
stages 
can be drawn from these correlations, with mergers becoming significantly
less concentrated, more asymmetric, and more clumpy than pre-mergers,
then progressing back toward the normal galaxy parameter space, where
they become merger remnants, which are more concentrated and
slightly more asymmetric and clumpy than pre-mergers.

We find no significant difference in the CAS parameters for 
early-type galaxies (E--S0) at rest-frame wavelengths longward of the 
Balmer break. We cannot measure the wavelength dependence of the CAS
parameters of early-type galaxies shortward of the Balmer break with
great confidence, because these red galaxies have very low S/N in 
the GALEX and HST F255W and F300W filters, and because of the lack 
of good statistics in the far and mid-UV. 
However, we find that later type
galaxies (S--Im and peculiar/mergers) observed at shorter wavelengths 
appear to be less concentrated, more asymmetric, and more clumpy
than they would appear to be at longer wavelengths. Therefore, there
is little to no morphological k-correction for the CAS parameters of
early-type galaxies (E--S0) in the optical, but a more significant
correction for all other later galaxy types at all wavelengths between 
the far-UV and near-IR. 

These nearby-galaxy results can be used as a benchmark for interpreting
the CAS morphological parameters of high redshift galaxies observed in
the rest-frame UV. The morphological k-correction derived here will help 
us to determine how much of the difference we see in the morphologies of 
high redshift galaxies is purely due to band-pass shifting effects, and a 
comparison of the structural CAS parameters between low and high redshift 
galaxies will provide some understanding of galaxy evolution. We will 
address these aspects of the high redshift Universe in a future paper 
(Conselice, et al.  2007, in preparation).

The current study shows that one cardinal improvement to this work would
be to obtain a much larger sample in the mid-UV at $\sim 100 \times$ the 
sensitivity, which WFC3 will be able to do after SM4 is launched. Such
studies will be critical to help interpret the 10's of millions of
distant galaxy images JWST will observe after 2013 in the rest-frame UV, 
at $\lambda \gtrsim 1\micron$ and z $\gtrsim 2$. 

\acknowledgments

Based on observations with the VATT: the Alice P. Lennon Telescope and the 
Thomas J. Bannan Astrophysics Facility.
This research was partially funded by NASA grants GO-8645.01-A, 
GO-9124.01-A and GO-9824.01-A, awarded by STScI,
which is operated by AURA for NASA under contract NAS 5-26555.
Additional funding was provided by GALEX grant \#GALEXGI04-0000-0036, 
and the NASA Space Grant Graduate Fellowship 
at ASU. We wish to thank the entire staff of Steward Observatory and
the Vatican Advanced Technology Telescope for all of their
help and support on this project. We especially thank Drs. Richard
Boyle, Matt Nelson, and Chris Corbally for offering so much of their
time and expertise during our VATT observing runs. 
CJC thanks the National Science Foundation for
support through an Astronomy and Astrophysics Fellowship and PPARC for
support at the University of Nottingham. This research has made use of the 
NASA/IPAC Extragalactic Database (NED) which is operated by the Jet 
Propulsion Laboratory, California Institute of Technology, under contract 
with the National Aeronautics and Space Administration.
This research has also made use of NASA's Astrophysics Data System.

\end{document}